\documentclass[twocolumn, times]{aastex61}
\usepackage{color}
\usepackage{captcont}
\usepackage[bookmarks=false]{hyperref}
\hypersetup{
  colorlinks= true, 
  urlcolor  = black, 
  linkcolor = red, 
  citecolor = blue 
} 

\shorttitle{ Faint submm galaxies}
\shortauthors{Patil et al.}

\begin{document}

\title{Multi-band Optical and Near-infrared Properties of Faint Submillimeter Galaxies with Serendipitous ALMA Detections}

\correspondingauthor{Pallavi Patil}
\email{pp3uq@virginia.edu}

\author[0000-0002-0786-7307]{Pallavi Patil} \altaffiliation{Reber Pre-Doctoral Fellow}
\affil{Department of Astronomy, University of Virginia, 530 McCormick Road, Charlottesville, VA 22903, USA}
\affiliation{National Radio Astronomy Observatory, 520 Edgemont Road, Charlottesville, VA 22903, USA}

\author{Kristina Nyland}
\affiliation{National Radio Astronomy Observatory, 520 Edgemont Road, Charlottesville, VA 22903, USA}

\author{Mark Lacy}
\affiliation{National Radio Astronomy Observatory, 520 Edgemont Road, Charlottesville, VA 22903, USA}

\author{Duncan Farrah}
\affiliation{Department of Physics and Astronomy, University of Hawaii, 2505 Correa Road, Honolulu, HI 96822, USA}
\affiliation{Institute for Astronomy, 2680 Woodlawn Drive, University of Hawaii, Honolulu, HI 96822, USA}

\author{Jos\'{e} Afonso}
\affil{Instituto de Astrof\'{i}sica e Ci\^{e}ncias do Espa\c co, Universidade de Lisboa, OAL, Tapada da Ajuda, PT1349-018 Lisboa, Portugal}
\affiliation{Departamento de F\'{i}sica, Faculdade de Ci\^{e}ncias, Universidade de Lisboa, Edif\'{i}cio C8, Campo Grande, PT1749-016 Lisbon, Portugal}

\author{Wayne Barkhouse}
\affiliation{Department of Physics and Astrophysics, University of North Dakota, Grand Forks, ND 58202-7129, USA}

\author{Jason Surace}
\affiliation{California Institute of Technology, 1200 E. California Boulevard, Pasadena, CA 91125, USA}



\begin{abstract}

We present a catalog of 26 faint submillimeter galaxies (SMGs) in the XMM-LSS field identified by cross-matching serendipitously detected sources in archival ALMA Band 6 and 7 data with multi-band near-infrared (NIR) and optical data from the \textit{Spitzer} Extragalactic Representative Volume Survey, the VISTA Deep Extragalactic Survey, the Canada-France-Hawaii Telescope Legacy Large Survey, and the Hyper Suprime-Cam Subaru Strategic Program. 
Of the 26 SMGs in our sample, 15 are identified here for the first time. The majority of the sources in our sample (16/26) have faint submm fluxes ($0.1\,{\rm mJy} < S_{\rm 1\,mm} < 1\,{\rm mJy}$).  
In addition to the 26 SMGs with multi-band optical and NIR detections, there are 60 highly-reliable ($>5\sigma$) ALMA sources with no counterpart in any other band down to an IRAC [4.5] $AB$ magnitude of $\approx 23.7$. To further characterize the 26 galaxies with both ALMA and optical/NIR counterparts, we provide 13-band forced photometry for the entire catalog using the {\it Tractor} and calculate photometric redshifts and rest-frame colors.  The median redshift of our sample is $\langle z \rangle = 2.66$. We find that our sample galaxies have bluer colors compared to bright SMGs, and the UVJ color plot indicates that their colors are consistent with main sequence star-forming galaxies. Our results provide new insights into the nature of the faint population of SMGs, and also highlight opportunities for galaxy evolution studies based on archival ALMA data.

\end{abstract}

\keywords{galaxies:evolution--galaxies:star formation--submillimeter:galaxies--galaxies:photometry--galaxies:high-redshift}


\section{Introduction} \label{sec:intro}

Dusty Star-Forming Galaxies (DSFGs) are a class of galaxies enshrouded in dust that have star formation rates of at least a few tens of solar masses per year \citep{casey2014dusty}. A characteristic feature of this class is the bright dust emission at Far-infrared (FIR) wavelengths, which is the re-radiated optical/ultra-violet light from star-forming regions. Their discovery became possible by the advances in IR instruments during the 1980s-1990s, and the {\em Infra-Red Astronomical Satellite (IRAS)} allowed a large number of detections of Luminous and Ultraluminous Infrared Galaxies (LIRGs and ULIRGs). Furthermore, the Cosmic Background Explorer (COBE) satellite was the first to measure the Cosmic Infrared Background (CIB) light and establish the overall importance of the DFSG population. The results from COBE showed that the energy density of the CIB emission is comparable to the optical and UV background light \citep{dole+2006, Hauser+1998}. Studies in the late 1990s (\citealt{Smail+1997, Barger+1998, Hughes+1998,Blain2002a}, and references therein) identified the galaxies responsible for the CIB submm/mm emission. These galaxies, popularly known as Submillimeter Galaxies (SMG; \citealt{Blain2002a}), belong to the more general class of DFSGs.


SMGs are extremely luminous galaxies ($L_{IR} >10^{12} L_\odot$) with star formation rates up to 1000$\,M_\odot$ yr$^{-1}$ \citep{Ivison+1998,Smail+2002,Chapman+2004,Barger+2012,Swinbank+2013,Simpson+2017,Mich+2017}. They are most likely undergoing a merger \citep{Engel+2010,Ivison+2012,AlZa+2012,Fu+2013,Chen+2015,Oteo+2016}, and have a median redshift, $\langle z \rangle \sim$ 2-3 \citep{Chapman+2002,Chapman+2005,Wardlow+2011,Simpson+2014,Miett+2015,Chen+2016}. 
Although early studies of SMGs only utilized observation windows at 850$\,\mu$m and 1.1$\,$mm due to atmospheric transmission, the negative $k$-correction helped to probe SMGs up to high redshift ($z \sim 5$). For a given luminosity, the dimming of the flux with increasing redshift is balanced by the shifting of the peak of the Spectral Energy Distribution (SED) into the observing window (\citealt{Franceschini+1991, Blain2002a}; See Fig.~4 in \citet{Blain2002a}). Therefore, the flux remains approximately constant for redshifts up to $z \sim 8$.

Single-dish instruments such as the Submillimetre Common User Bolometer Array (SCUBA; e.g. \citealt{Smail+1997, Hughes+1998, Barger+1998}) on the James Clark Maxwell Telescope (JCMT),  AzTEC \citep{Ezawa+2004, Perera+2008, Austermann+2010, Aretxaga+2011, Scott+2010, Scott+2012} on the Atacama Submillimeter Telescope Experiment (ASTE), the Large Apex Bolometer Camera (LABOCA) on the Atacama Pathfinder Experiment (APEX) \citep{Siringo+2009,Weib+2009}, Bolocam on the Caltech Submillimeter Observatory (CSO) \citep{Laurent+2005}, the  Max Planck Millimeter Bolometer (MAMBO; \citealt{Greve+2004,Bertoldi+2007}) on the IRAM 30-meter telescope, and SCUBA-2 on JCMT \citep{Chen+2013b,Casey+2013,Geach+2013,Geach+2017} identified a significant number of SMGs. However, large beam sizes (15$^{\prime \prime}$ for SCUBA at 850 $\,\mu$m and 11$^{\prime \prime}$ for MAMBO-1 at 1.2$\,$mm) prevented accurate multi-wavelength counterpart identification. A few indirect techniques were used to identify multi-wavelength counterparts, such as using the radio-FIR correlation to find targets in radio interferometric observations \citep{Ivison+2007}, 24$\,\mu$m MIPS observations from {\it Spitzer} (e.g. \citealt{Pope+2006}), and assigning probabilities to the possible optical/NIR counterparts (e.g. \citealt{Ivison+2002, Dunlop+2002,  Chapin+2011,Smith+2011,Kim+2012,Alberts+2013}, \citealt{Biggs+2011}). These techniques have limitations as not all counterparts can be accurately identified due to large beam sizes of single dish telescopes. 
Therefore, interferometric observation is required to obtain a bias-free sample of submm sources with sub-arcsecond positional accuracy.
%
%

Number count studies revealed that a significant population was missed by the single dish as well as space-based observations at submm and mm wavelengths (\citealt{Lagache+2005}).
%
%
On the other hand, most of the population contributing to the CIB at wavelengths less that 200$\,\mu$m was already identified, and found to reside at $z \sim 1$ \citep{Viero+2013}. Number counts as well as models, showed that the redshift of the dominant contributor to the CIB increases with increasing wavelength \citep{Lagache+2005}. This indicates the need for high sensitivity and finer spatial resolution surveys. During the pre-Atacama Large Millimeter Array (ALMA) era, 1.1$\,$mm surveys were mainly conducted with AzTEC \citep{Perera+2008, Scott+2012, Hatuskade+2011, Austermann+2010,Scott+2010,Aretxaga+2011} and about $10-20$\% of the CIB was resolved \citep{Hatuskade+2011, Scott+2010}. In contrast, 850/870$\,\mu$m surveys using SCUBA, the  LABOCA has resolved up to 50$\%$ of the CIB \citep{Blain+1991,Coppin+2006,Weib+2009}\footnote{We have provided only a selected number of references here. Refer to Section~3 in \citet{casey2014dusty} for a complete list.}. The high confusion limit limited the detection threshold to $\sim$ 1-2$\,$mJy at $850\,\mu$m. Studies of gravitationally-lensed SMGs allowed to probe the faint population \citep{Hsu+2016,Chen+2013a, Chen+2013b,Smail+1997, Cowie+2002, Knudsen+2008,Johansson+2011, Chen+2014}. However, these studies were limited by small number statistics and uncertainties in the lensing models for the clusters. In summary, all of the previous results have shown that the major contributors of the CIB at 850$\,\mu$m and 1.1$\,$mm have flux densities fainter than 1$\,$mJy. Such galaxies would correspond to normal galaxies, or LIRGs (with luminosities $L < 10^{12}\,L_\odot$).

It is now possible to study this fainter population, the so-called faint SMGs, as ALMA provides sub-arcsecond resolution and high sensitivity at submm and mm wavelengths. However, the small field of view of ALMA makes large surveys challenging. Several groups have tried different approaches to optimize ALMA's resources and search for the faint population. One approach is to look for serendipitous detections using archival observations obtained for other scientific goals \citep{Hatsukade+2013, Ono+2014, Carniani+2015,fujimoto+2016, Oteo+2016}. 
Another approach is to observe a contiguous field using ALMA \citep{Dunlop+2017,Kohno+2016, Gonzalez+2017, Franco+2018}.

A third approach made use of the available deep optical-NIR surveys  to develop a triple color selection technique for identification faint galaxies \citep{Chen+2016}. These pilot studies are attempting to address the contribution of faint SMGs to the CIB, their multi-wavelength counterparts, and their role in shaping galaxy formation. The number count studies revealed that faint SMGs contribute significantly to the extragalactic background light (EBL) at 1.1$\,$mm. \cite{fujimoto+2016} found that the contribution of faint SMGs (0.02$\,$mJy $<$ $S_{1.2\,mm}$ $< $ 1$\,$mJy) can account for all of the CIB at 1.2$\,$mm
If we assume a median redshift, $z \sim$ 2, the IR luminosities of faint SMGs are expected to be in the range $10^{11-12}\,$L$_\odot$ (\citealt{Chen+2016}). Therefore, this population bridges the gap between extreme star-forming galaxies (bright SMGs) and optical-color selected galaxies with moderate star-formation rates, e.g., Lyman Break Galaxies (LBGs)\footnote{See \citet{Steidel+1996} for more on the properties of LBGs.} and star forming BzK galaxies (sBzKs)\footnote{ See \citet{Daddi+2004} for the formal definition of SBzKs.} \citep{Chen+2016}. 


Based on optical-NIR color-color plots, the faint SMGs from \citet{fujimoto+2016} were found to represent LBG/sBzK galaxies. However, these studies suffer from small number statistics, and very little is known about this newly discovered population. Robust estimates of demographics, such as number counts, the redshift distribution, the contribution to the cosmic star formation rate density (SFRD), the nature of their multi-wavelength counterparts, and the star formation rate distribution are some of the key issues that need to be addressed.

In this paper, we study faint SMGs in the XMM Large Scale Structure (XMM-LSS) field with multi-band optical and infrared survey data that have serendipitous submm counterparts identified in archival ALMA observations. Many of the sources identified in our sample are faint SMGs, with $\sim\,1\, $mm fluxes below 1$\,$mJy. We investigate the properties of this cosmologically important galaxy population by performing multi-band forced photometry to obtain photometric redshifts and place constraints on star formation.  Our study also highlights the growing opportunities for probing high-redshift galaxy properties and gaining new insights on cosmic assembly by mining the ALMA archive.

In Section~\ref{sec:obs_and_data}, we describe our sample selection procedure.  Details on our reduction of the archival ALMA data, source finding strategy, multi-band photometric catalog construction, and photometric redshift determination are given in Section~\ref{sec:analysis}.  We discuss the multi-wavelength source properties of our sample in Section~\ref{sec:discussion} and provide a summary of our results in Section~\ref{sec:summary}.  Throughout this study we adopt a $\Lambda$CDM cosmology with $\Omega_{\mathrm{M}}$ = 0.3, $\Omega_{\Lambda}$  = 0.7, and H$_0$ = 70 km~s$^{-1}$ Mpc$^{-1}$.


\section{Sample}
\label{sec:obs_and_data}

\subsection{Optical and Infrared Data}
\label{sec:optical_ir_data}
Our sample is drawn from the XMM-LSS field. The XMM-LSS field includes abundant multi-band data at optical and infrared wavelengths from a variety of wide-field surveys.  Of particular importance is the availability of deep {\it Spitzer} IRAC observations at 3.6 and 4.5~$\mu$m from the {\it Spitzer} Extragalactic Representative Volume Survey (SERVS; \citealt{mauduit+12}) and DeepDrill (P.I. Mark Lacy). SERVS is a post-cryogenic IRAC survey of five well-studied astronomical deep fields with a depth of 2~$\mu$Jy and a total sky footprint of 18 deg$^2$.  The DeepDrill survey expands upon the sky coverage of SERVS and provides deep IRAC imaging in three of the four predefined Deep Drilling Fields for the Large Synoptic Survey Telescope over an area of 38.4~deg$^2$ ($\sim$1~Gpc$^3$ at $z > 2$).  

\begin{figure*}[t!]
\centering
\includegraphics[width=6in]
{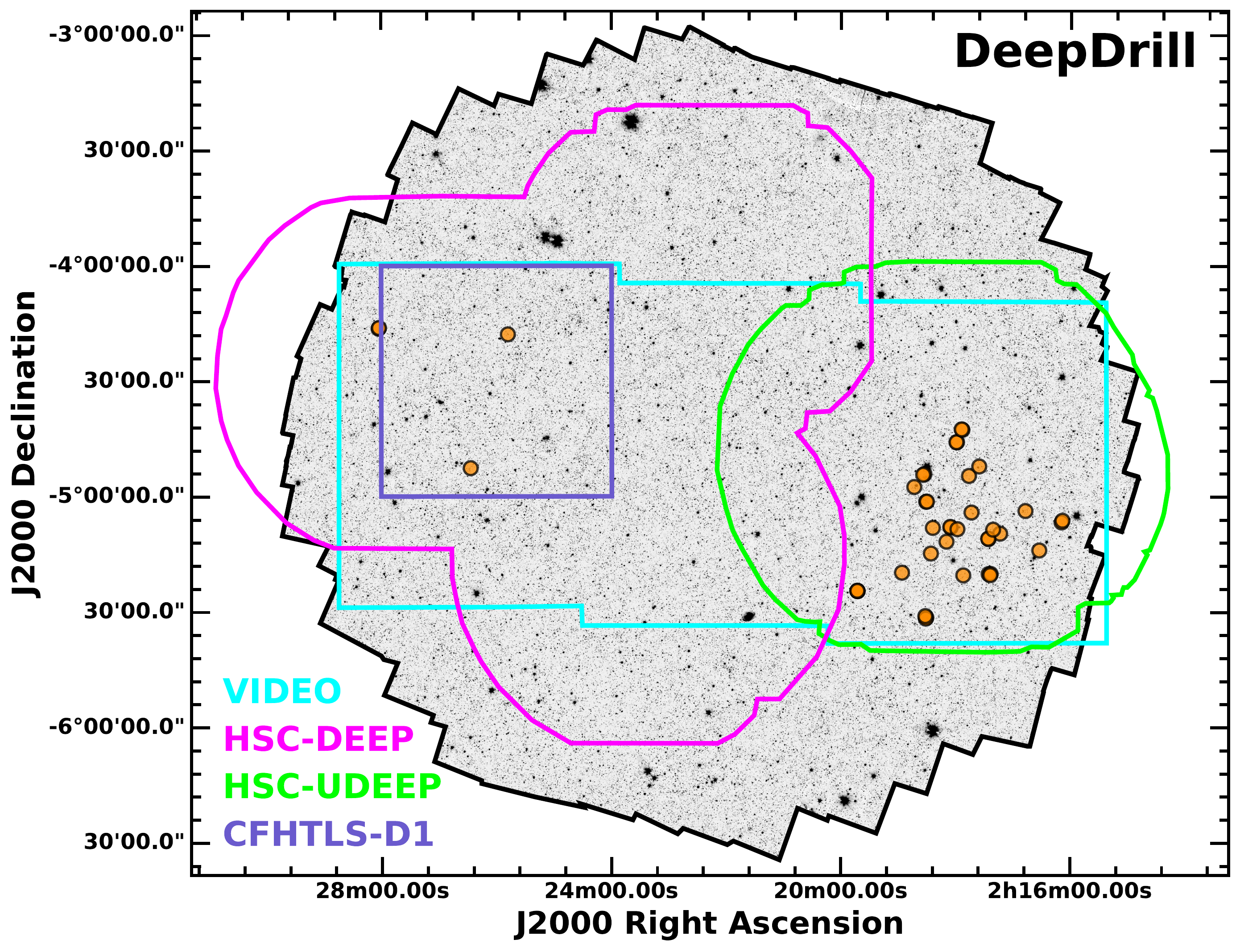}
\caption{The multi-band coverage by the surveys used in our study in XMM-LSS field. The grayscale image is the DeepDrill 3.6$\,\mu$m mosaic, the cyan region traces the VIDEO coverage, the green region shows the HSC ultradeep tier, the magenta region denotes the HSC deep tier, and the purple region shows the CFHT-LS coverage.  The orange circles indicate the ALMA sources with optical/NIR counterparts identified in this study.\\}
\label{fig:footprint}
\end{figure*}

Observations at 3.6 and 4.5~$\mu$m are crucial for detecting rest-frame optical emission from galaxies at $z > 4$, and, when combined with additional photometry at optical and NIR wavelengths, provide constraints on important galaxy properties such as photometric redshift. The SERVS and DeepDrill observations in the XMM-LSS field are complemented by additional NIR data from the ground-based VISTA Deep Extragalactic Survey (VIDEO; \citealt{jarvis+13}) in the $Z$, $Y$, $J$, $H$, and $Ks$ bands.  In the optical, wide-field data are available from the Canada-France-Hawaii Telescope Legacy Survey Wide field 1 (CFHTLS-W1; \citealt{gwyn+12}) and multiple tiers from the first data release of the Hyper Suprime-Cam Subaru Strategic Program (HSC; \citealt{aihara+17}).  We include the NIR data from SERVS/DeepDrill and VIDEO, as well as broad-band optical data from HSC in the $grizy$ filter set and data from CFHTLS-W1 in the $u$ band in our analysis.  Thus, we have a total of 13 bands available for determining photometric redshifts. We illustrate the sky coverage of these surveys in Figure~\ref{fig:footprint}. 


\subsection{Archival ALMA Data}
We mined the ALMA archive to search for continuum observations within the XMM-LSS field that were publicly available as of July 2017.  
We required the following criteria for selecting the archival ALMA data: 1) observations performed in Band~6 ($211-275$~GHz) or Band~7 ($275-370$~GHz), 2) a source integration time longer than 150 seconds to ensure sufficient sensitivity to the inherently faint SMG population , and 3) an angular resolution of $\theta_{\mathrm{FWHM}}$ $>$ 0.4$^{\prime \prime}$ to ensure adequate surface brightness sensitivity. After evaluating the central depths for all the available programs making the resolution cut, we found that our integration time criterion leads to a maximum 1$\sigma$ rms noise of 150~$\mu$Jy~beam$^{-1}$ in Band~6 and 300~$\mu$Jy~beam$^{-1}$ in Band 7, respectively. Given the nature of our source search, the source depth is variable \footnote{The source search depth depends on the integration time as well as the distance of the source from the ALMA pointing center.  As the noise increases towards the edge of the field of view due to the primary beam response, the sensitivity is variable within each pointing. Therefore, specifying the completeness limit would not be meaningful for our study, which aims to explore the properties of the faint SMG population rather perform a statistically complete analysis. }
and therefore, finding a complete census of faint SMG is not a primary goal of our search in this paper. We aim to find as many faint galaxies as we can to build up a large sample of the faint population having comprehensive multiwavelength coverage. Such a sample will allow us to conduct detailed studies which will help guide future statistical studies.

Our search identified 75 continuum maps from nine different projects \footnote{The nine public archival ALMA projects were undertaken for entirely different science goals. Brief descriptions of those goals are given in Appendix~\ref{sec:appB}.}, all but one observed in Band~6.  The typical RMS noise levels for Band~6 ($\sim$1.2~mm) and Band~7 (870~$\mu$m) are 15-140~$\mu$Jy~beam$^{-1}$ and 200-300~$\mu$Jy~beam$^{-1}$, respectively. Therefore, we note that the image depth is not uniform throughout our sample.  Table~\ref{tab:tab1} summarizes the list of ALMA pointings considered for our study.  The angular resolutions of the archival data ranged from 0.53$^{\prime \prime}$ to 1.46$^{\prime \prime}$. 
\section{Data Analysis}
\label{sec:analysis}

\subsection{ALMA}
\subsubsection{Calibration and Imaging}
We reduced the archival ALMA data using the Common Astronomy Software Applications (CASA; \citealt{McMullin+2007}) package. 
We ran the calibration scripts that are provided along with the raw data from the ALMA archive. The calibration scripts include a-priori flagging, bandpass calibration, flux calibration and gain calibration. The calibrated products were examined for further flagging in the uv-plane as well as image plane. We found that the provided script had flagged most of the bad data and very little additional flagging was required. 

We used the CLEAN task to form and deconvolve images with the recommended parameters provided by the ALMA observatory. Specifically, CLEAN was run in multi-frequency synthesis (MFS) mode with nterms = 1. 
The weighting was either Natural or Briggs weighting with a robust parameter of 0.5, and was determined on a case by case basis. 
Maps with bright targets were self calibrated (using one round of phase-only self calibration) and re-imaged.


\subsubsection{Source Extraction}
\label{sec:pybdsf}
The Python Blob Detector and Source Finder (PyBDSF) tool was used to extract sources from the ALMA maps \citep{pybdsf}. 
Continuum maps without primary beam corrections were used to search for sources. The algorithm looks for image pixels above a specified threshold (here we used threshpix = 3.0). Contiguous pixels above the threshold with a minimum size of 1/3 of the  
synthesized beam are formed into a single island. An island is considered a valid source if a single or multiple component Gaussian fit is successful. 

We have found that PyBDSF works well with the default parameters.  However, we set the pixel threshold parameter to 3.5$\sigma$ instead of 5$\sigma$, as the default threshold was too conservative to probe the fainter population. By lowering  the threshold, we are increasing the contamination, 
but prior source position information from 
the multi-wavelength optical/infrared data will eliminate most of the spurious sources. The preliminary catalog contains all extracted sources, including the science targets of the proposed observations.  We have also removed the sources lying at a distance larger than the radius where the primary beam sensitivity drops to 10\% of its maximum, and four strongly lensed galaxies that were the targets of some of the original observations.
Although PyBDSF provides estimates of several parameters; total and peak flux densities, convolved and deconvolved source sizes, and uncertainties on each parameter, we chose only to use it to identify ALMA source position. Those source positions were then 
cross-matched with the multi-band optical/infrared data (see Section \ref{tractor}).



In order to avoid known issues with flux overestimation of faint sources with PyBDSF \citep{hopkins+15}, we used the JMFIT task from the Astronomical Image Processing Software (AIPS) to measure source fluxes and their uncertainties.  For each ALMA source, we used JMFIT to fit a two-dimensional, single-component Gaussian at the position from our source extraction with PyBDSF.  All JMFIT measurements were based on the primary-beam-corrected ALMA images.  We have tabulated the ALMA fluxes of our sample sources in Table~\ref{tab:almaobs}.

\subsubsection{Detection Threshold}
Some spurious detections are likely to contaminate the source catalog, assuming pure Gaussian-like noise. Therefore, it is necessary to determine the signal-to-noise ratio (SNR) cutoff at which an optimal compromise is made between minimizing the number of spurious sources and maintaining a reasonable level of completeness for faint objects. 
One way to quantify the level of spurious source contamination is to perform a negative peak analysis 
(\citealt{fujimoto+2016,Hatsukade+2013,Ono+2014,Carniani+2015}).  
To accomplish this, we multiplied each ALMA image by -1 and ran PyBDSF using the same input parameters as the ones used in the original source extraction. 
We then plot separate SNR distributions for all sources extracted from the negative and positive peak analyses.  

\begin{figure*}

\includegraphics[width= \textwidth]{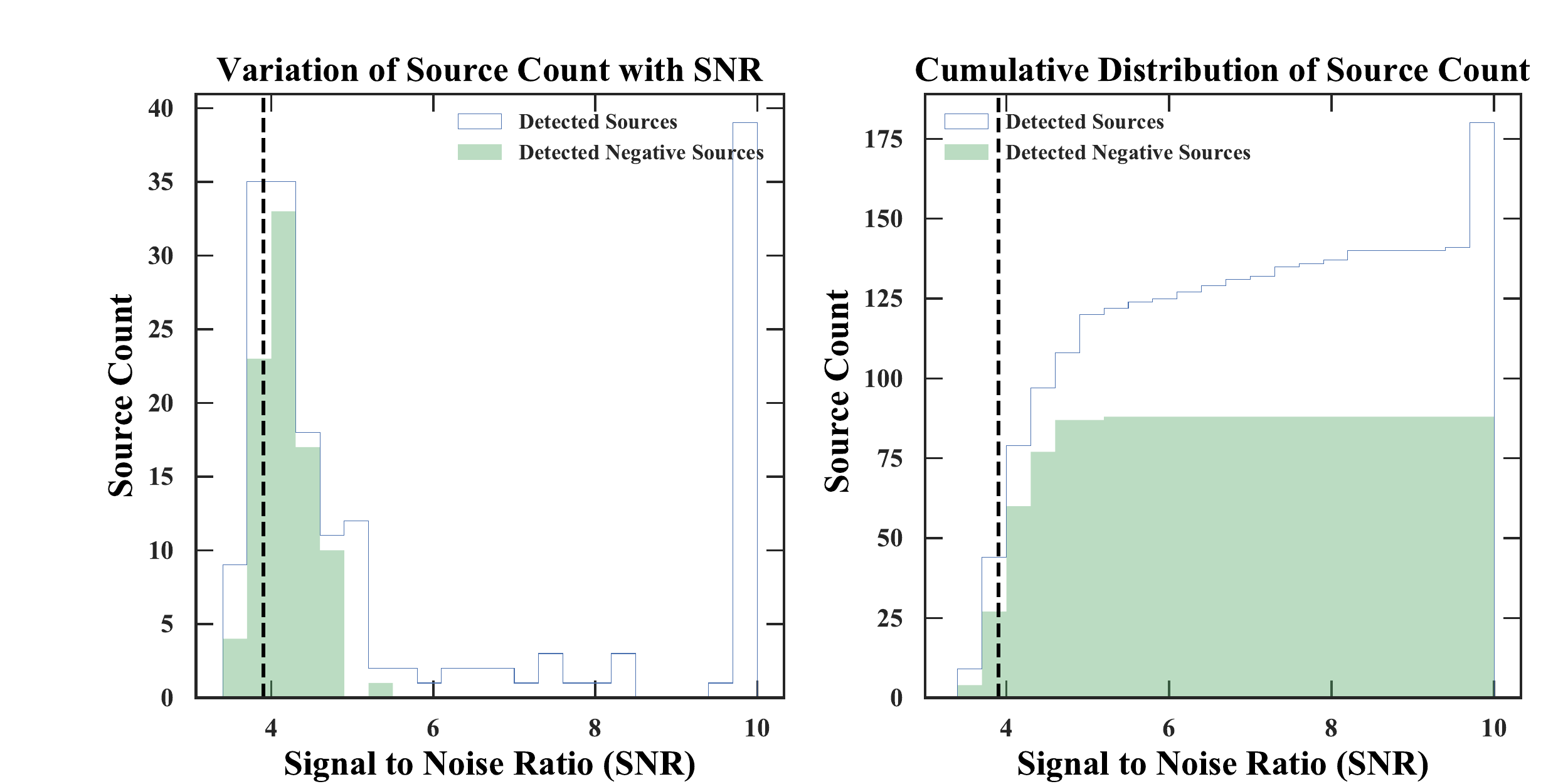}
\caption{Detection threshold for the archival ALMA images.  The left panel shows the differential source count and the right panel shows cumulative source count.  This analysis was used to select a detection threshold of 3.9$\sigma$ for the archival ALMA images.\\}
\label{fig:negpeak}
\end{figure*}

If a given map only contains noise and no real emission, then the total number of sources in the positive and negative maps would be approximately the same. This will result in a similar source distribution as a function of SNR. However, when real sources are present, we will start to see an excess of positive sources  over negative sources above a certain SNR value. The detection threshold for the source catalog can, then, be chosen at a certain value after which the number of positive detections are greater than negative ones.  
Figure \ref{fig:negpeak} shows the number of sources extracted from both the positive and negative maps from our work. 
Based on this figure, we selected a detection threshold of 3.9$\sigma$ for the archival ALMA data. Once this threshold was applied, our ALMA catalog was reduced to 176 objects. We then cross-matched this catalog with the optical-NIR photometry (see Section~\ref{tractor}) and found 26 faint SMGs with counterparts within a search radius of 1$\arcsec$.

In order to check the fidelity of our source selection, we performed the same counterpart matching steps for our negative ALMA source catalog and found 7 out of 88 sources with optical-NIR counterparts. Whereas, we found optical-NIR cross-matches for 26 out of 176 sources in the positive source search. Thus, the combination of our detection threshold in the ALMA data and an optical-NIR counterpart leads to a significantly greater level of fidelity compared to the level (50\%) that the 3.9$\sigma$ ALMA detection threshold alone would provide.

In extracting sources from the negative maps, we did find a few targets with multi-Gaussian structures. As we do not expect to see any complex sources, this could be due to image artifacts. Therefore, we excluded sources with such structures from our catalog. 
We note that we have not estimated the formal completeness of our ALMA catalog, nor are we conducting a number count analysis. Since we are limited by the availability of data in the archive, our sample will naturally be incomplete. In this paper, we focus on investigating the multi-wavelength properties of our sample of faint SMGs.



\subsubsection{Comparison with Previous Studies}

A few of the ALMA pointings in our sample and that of \citet{fujimoto+2016} overlap, which provides us with an opportunity to compare methodologies, source counts, and source fluxes in the respective samples. In total, 20 ALMA pointings from three different archival programs (ALMA \#2011.0.00115.S, \#2011.0.00648.S, \#2012.1.00934.S) are common with \citet{fujimoto+2016}. The source extraction parameters for our study are different from \citet{fujimoto+2016}. The authors have selected sources above SNR of 3.4$\sigma$ within a search radius having primary beam sensitivity of 50\% or larger. Whereas, we used 3.9$\sigma$ as an SNR threshold and a search radius having primary beam sensitivity greater than 10\% of its maximum. In our source extraction, we found 26 sources above 3.9$\sigma$ within those 20 pointings. And \citet{fujimoto+2016} found  14 sources above 3.4$\sigma$, out of which 6 sources are above 3.9$\sigma$. Those six sources are also detected in our sample. Therefore, source counts of both the studies are consistent when same source extraction parameters are selected. Furthermore, the estimated flux densities also agree within the quoted uncertainties for those six sources.

\subsection{Multi-band Forced Photometry}\label{tractor}
In order to proceed with our photometric redshift and SED analysis, construction of a robust multi-band photometric source catalog is necessary.  However, the difference in angular resolution between the {\it Spitzer} IRAC ($\sim$2$^{\prime \prime}$) and ground-based optical/NIR survey data ($\lesssim$1$^{\prime \prime}$), coupled with the crowded nature of these observations, make the IRAC data prone to issues with source blending.  This is problematic for accurate source cross-identification between bands and reliable multi-band photometry.  

Recently, \citet{nyland+17} demonstrated a means of mitigating many of the issues inherent to mixed-resolution optical/NIR datasets using a ``forced photometry'' approach based on the {\it Tractor} imaging modeling code \citep{lang+16}.  This code uses prior information on source position and surface brightness profile from a high-resolution, ``fiducial'' band, along with image calibration parameters including the point spread function (PSF), to model the source flux in lower-resolution bands.  After applying the {\it Tractor} to a one square degree test region of the XMM-LSS field, \citet{nyland+17} found a number of improvements in the resulting multi-band forced photometry compared to traditional position-matched source catalogs.  In particular, they found that the {\it Tractor} forced photometry decreased susceptibility to blending issues, led to more reliable source cross-matching between bands, identified a larger number of candidate high-redshift ($z > 5$) objects, and produced more accurate photometric redshifts when compared to available spectroscopic redshift data.  

We have adopted a strategy similar to that described in \citet{nyland+17} in constructing the optical/NIR source catalog used for determining photometric redshifts of the ALMA sources in our study.  This strategy requires an initial, position-matched input catalog that is constructed by cross-matching the positions of the ALMA sources with positions from VIDEO\footnote{VIDEO source catalogs and images were obtained from the fifth data release available at http://horus.roe.ac.uk/vsa/.} using a search radius of 1$^{\prime \prime}$.  Thus, each source in this ``VIDEO-selected'' input catalog has a detection in the VIDEO catalog in at least one band.  

In addition, a ``fiducial'' VIDEO band is selected for each source that is used for determining the source surface brightness profile model to be applied during the {\it Tractor} forced photometry.  We preferentially select the VIDEO $Ks$-band to be the fiducial band, but if the source is not detected at $Ks$-band, we set the fiducial VIDEO band to the next closest filter in wavelength to the IRAC 3.6~$\mu$m band that has a detection reported in the VIDEO source catalog.  We note that of 26 ALMA sources with optical-NIR counterparts in our catalog, 23 reside in the UltraDeep tier of HSC (5$\sigma$ $i$-band depth $\sim$ 27.2 mag) and 3 sources reside in the Deep HSC tier (5$\sigma$ $i$-band depth $\sim$26.5 mag).  
%
%

For each source in our VIDEO-selected input catalog, we extracted a cutout of the image in each of the 13 bands in our analysis with a half-width of 10$^{\prime \prime}$.  To account for spatial variations in the image properties, we measured the sky (RMS) noise and the median background sky level in each image cutout using iterative sigma clipping with Photutils\footnote{
\href{http://photutils.readthedocs.io/en/stable/}{http://photutils.readthedocs.io/en/stable/}}.  We then fit the source fluxes using the {\it Tractor}, which convolves the source surface brightness profile model with the image PSF and uses a maximum likelihood method to optimize the flux of each source in each band, holding the source position, shape, and image calibration properties fixed.   
We provide the resulting {\it Tractor} catalog of the forced photometry in Table~\ref{tab:tractor}. We also compare the photometry using the {\it Tractor} with other surveys covering the same field to check consistency of our results. Here we utilize publicly available data release 8 of the UKIDSS-UDS survey. We refer our readers to Appendix~\ref{sec:pcheck} for  the detailed discussion.



\subsection{Photometric Redshifts}\label{sec:photoz}

Photometric redshifts were estimated using the Easy and Accurate $z_{\mathrm phot}$ from Yale (EAZY; \citealt{Brammer+2008, Brammer+2011}) software. EAZY performs least square fitting with a linear combination of minimal template sets that can accommodate most of the variations in galaxy properties up to high redshift. 

We use a default set SED templates available in EAZY. These templates adequately span the wavelength range to cover the 13-bands used in our analysis (4.5$\,\mu$m to $u$ band). EAZY provides a full probability distribution for the redshift values in the range $0 < z_{\mathrm{phot}} < 6$ . We select the photometric redshift of a given source at the value with the highest probability ($z_{\mathrm{peak}}$). The confidence intervals are selected such that the integrated probability distribution is equal to 95\%, corresponding to the EAZY output parameters  \textit{l}95 and \textit{u}95). Table~\ref{tab:almaobs} provides the photometric redshifts  along with the 95\% confidence limits of our sample.

\begin{figure*}
\centering
\includegraphics[width=0.9\textwidth]{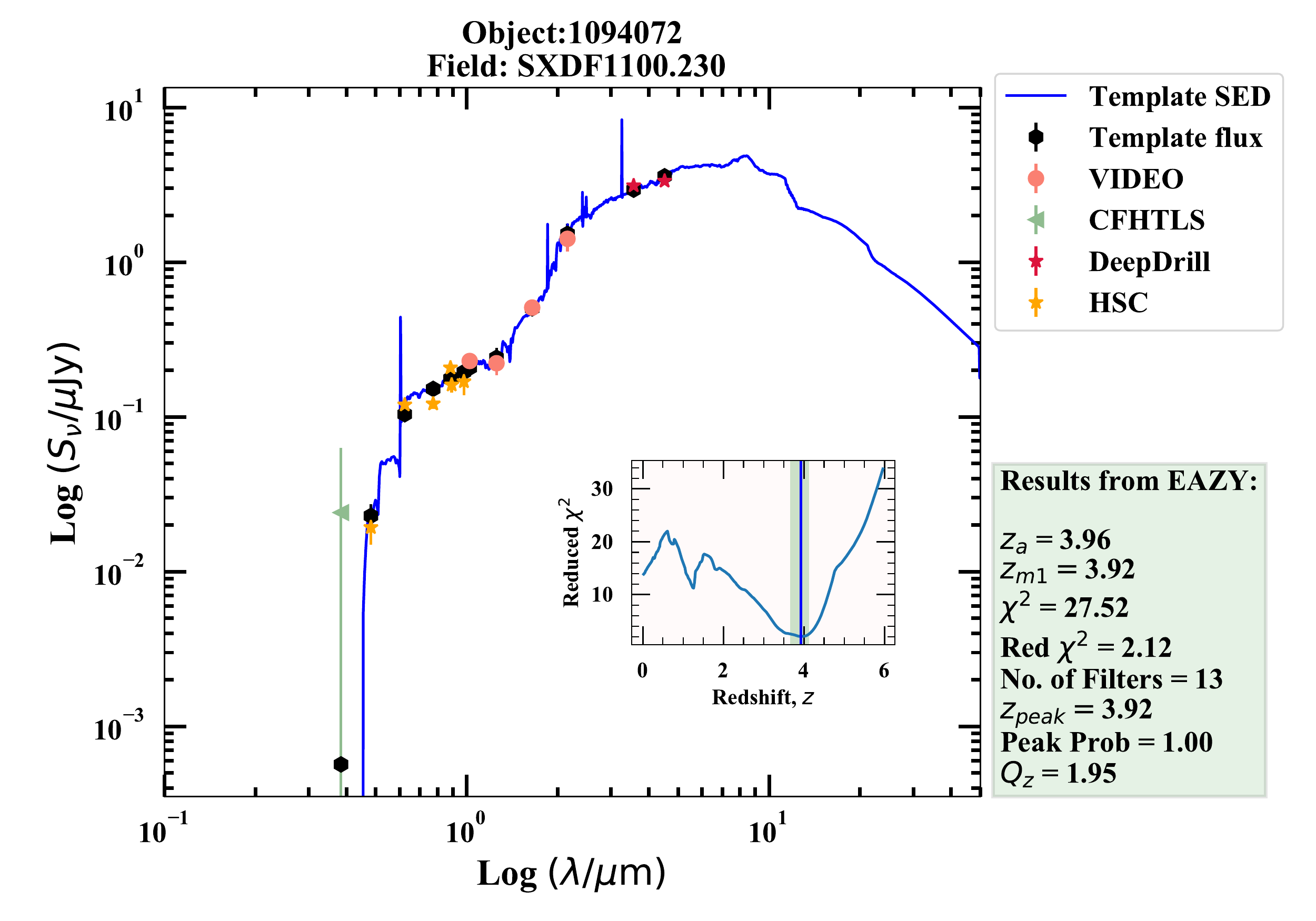}
\caption{An example of our photometric redshift fitting using EAZY for one of the sources in our sample. The main plot is the observed galaxy SED. Different colored symbols correspond to the observed 13-band photometry obtained using the {\it Tractor} image modeling code (Section~\ref{tractor}). Red stars highlight {\it Spitzer} DeepDrill measurements at 3.6 and 4.5~$\mu$m, orange circles denote measurements from the VIDEO survey, the green triangle represents the optical $u$-band data from CFHTLS, and yellow stars mark the optical data points from HSC. The fitted SED template is plotted in blue and the estimated template flux for each filter is shown by the black filled circles. The green box on the right lists key output parameters from EAZY. The inset figure shows the $\chi^{2}$ distribution of the fit at each redshift bin. The peak of the redshift distribution after applying the prior is shown by the  blue vertical line and the light green regions are the 95\% confidence intervals of the fit.\\}
\label{fig:eazyfit}
\end{figure*}

We illustrate the EAZY photometric redshift fitting results for our sample galaxies in Figure~\ref{fig:eazyfit}. As shown in the green-shaded box in this figure, EAZY provides several diagnostic parameters to quantify the quality of the fit. The fitting results for the entire sample are shown in the appendix Figure~\ref{fig:eazy_all}. We emphasize that our photometric redshifts are based on multi-band forced photometry using the \textit{Tractor}, a technique that has demonstrated improved photometric redshift accuracy compared to the use of position-matched multi-band photometric catalogs \citep{nyland+17}.  However, given the inherently dusty nature of our sources that may cause deviations in their SEDs that are not well represented by the templates considered in this study, further verification of their redshifts will require a more in-depth SED analysis (to be presented in a forthcoming study) as well as future spectroscopic observations.

The rest frame colors are also evaluated using best fit SED template in EAZY. To estimate colors, we have used filters $U$, $V$, and $J$\footnote{$U$ and $V$ are standard Bessel filters and the $J$ filter follows the definition of Mauna Kea Consortium defined in \citet{J_filt_02}.}. EAZY provides interpolated color indices for each filter of the rest-frame color, such that the colors can be calculated using the following formula; $U-V = -2.5\times \log(L_U/L_V)$, where $L_U$ and $L_V$ are the 
calculated $U$ and $V$-band luminosities, respectively.





\subsection{The Final Catalog }

We present a catalog of 26 galaxies obtained by cross-matching the ALMA source catalog with optical-NIR observations 
(Section~\ref{sec:optical_ir_data}). 
The catalog contains 15 new, serendipitously discovered ALMA detections. From the remaining 11 known sources, nine 
were detected in the previous studies, and three correspond to a single bright {\it Herschel} source \citep{Bussmann+2015}. We explain the properties of all previous detections later in this section. Figure~\ref{fig:cutouts} shows the four-band (ALMA Band 6 or 7, {\it Spitzer} 3.6 and 4.5$\,\mu$m, and VIDEO $K_s$) snapshots of the entire catalog. 
Table~\ref{tab:almaobs} summarizes the ALMA properties of our sample sources, our photometric redshifts, 
and any previously published redshift information. 
Throughout this paper, we will be identifying sources by their IAU names given in column 1 of  Table~\ref{tab:almaobs}. 


Nearly half of our targets (11/26)
were 
previously detected in the SIRTF Wide-area Infra Red Extragalactic Survey (SWIRE; \citealt{Lonsdale+2003}) and have published photometric redshifts using the five-band SWIRE  observations \citep{swire+z, Robinson+2013}. Another study that included a large number of our sample galaxies (13 out of 26) is that of \citet{Williams+2009}. These authors presented a $K$-band-selected galaxy catalog combining optical-mid IR photometry from SWIRE, the Ultra Deep Survey (UDS) tier of the UKIRT Infrared Deep Sky Survey (UKIDSS; \citealt{Lawrence+2007}), and the Subaru-XMM Deep Survey (SXDS; \citealt{Sekiguchi+2004, Furusawa+2008}). They have used EAZY to estimate photometric redshifts based on \textit{BRi$^{\prime}$z$^{\prime}$JK}[3.6$\,\mu$m][4.5$\,\mu$m] photometry. Our photometric redshifts are in good agreement with previously published values (when available) within the margin of the uncertainties. We emphasize that the 13-band forced photometry presented here provides redshift estimates that are robust against the effects of source blending in the IRAC bands and utilize a large number of filters compared to the previous studies.

\textbf{}


\section{Discussion}
\label{sec:discussion}

\subsection{Flux and Redshift Distribution}

\begin{figure}
\includegraphics[width=\linewidth]{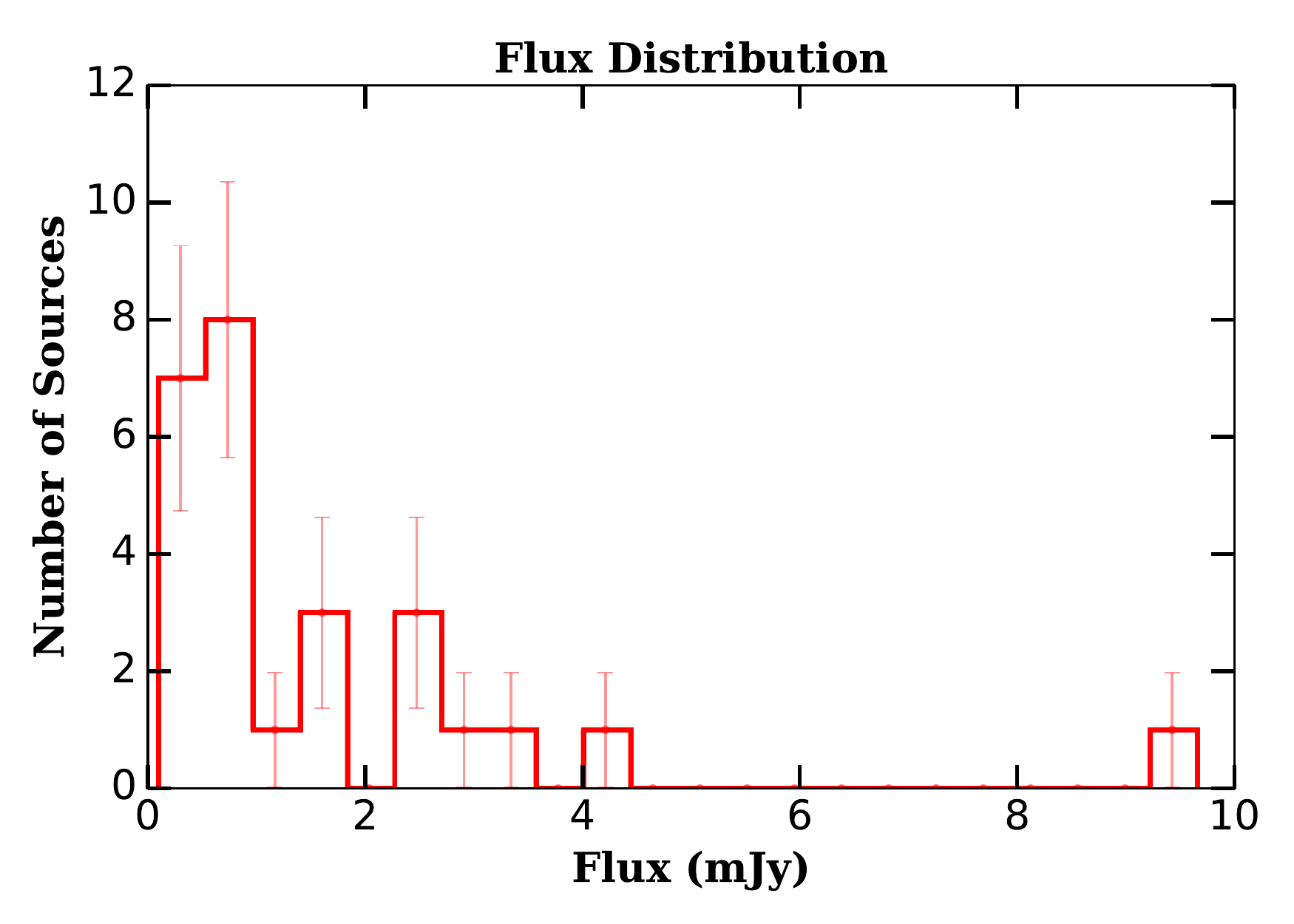}
\caption{ The flux distribution of our sample. The red line at the center of each bin is the binomial uncertainty. 
There are 15/26 galaxies (57$\%$) with fluxes fainter than 1~mJy. Except for one very bright galaxy ($S \sim 9.5\,$mJy), the rest of the bright sample has fluxes ranging from 1 to 5~mJy.\\}
\label{fig:flux_dist}
\end{figure}

\begin{figure}
\includegraphics[width=\linewidth]{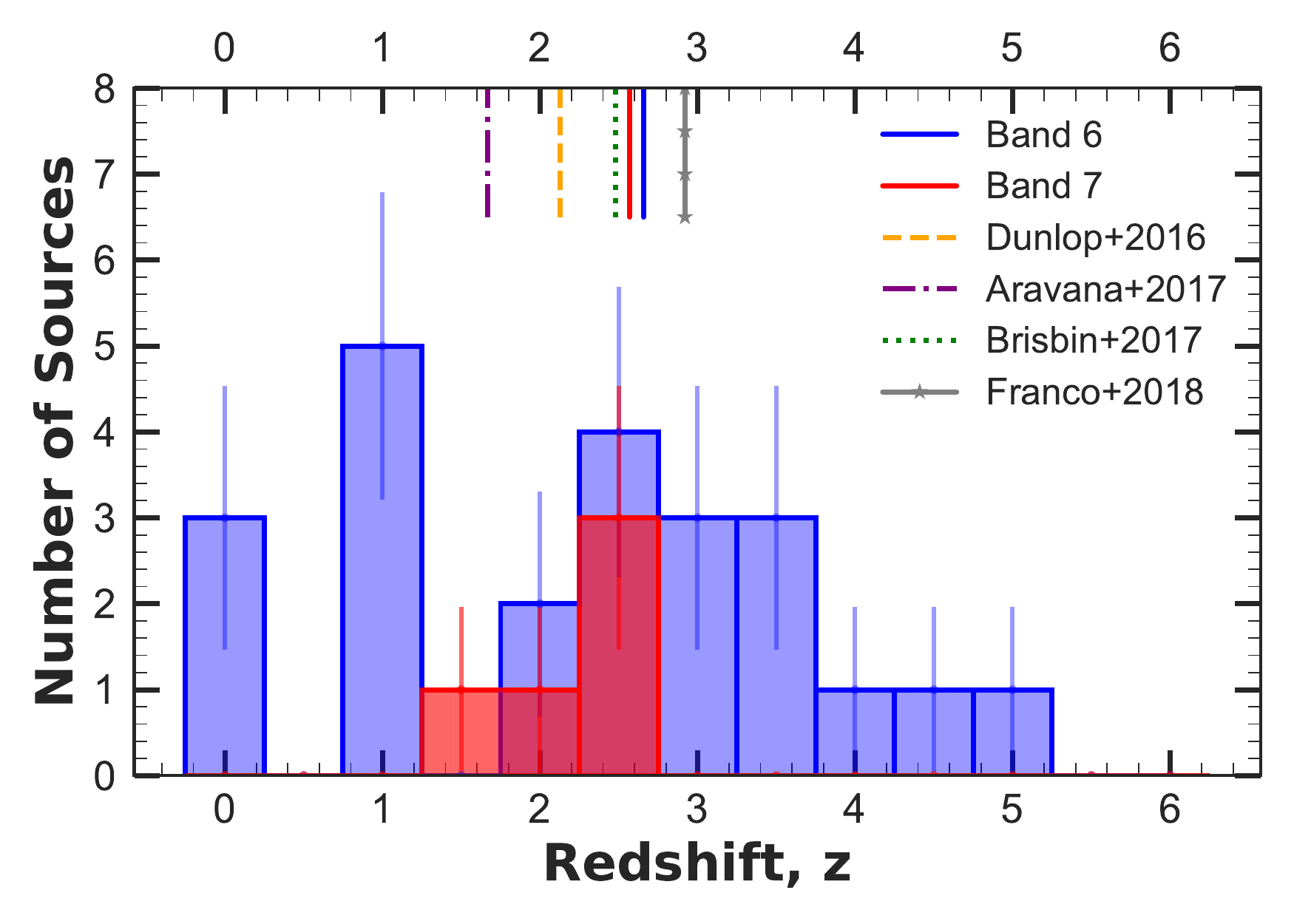}
\caption{ The redshift distribution of the faint SMGs. The photometric redshifts are evaluated using EAZY (Section~\ref{sec:photoz}). We show separate histograms for the ALMA frequency bands, 6 and 7. A line at the center of the bin shows the binomial uncertainty in each bin. The vertical lines at the top are the median redshifts for our sample as well as other faint SMG samples. Blue and red solid lines show the median values for the Band 6 and 7, respectively. Other studies included in the plot are: \citet{Aravena+2016} -- Red dash-dotted line; \citet{Dunlop+2017} -- Orange dashed line; \citet{Brisbin+2017} -- Green dotted line; \citet{Franco+2018} -- Silver solid line with star symbols. 
The average depth for our Band 6 sample is 110$\,\mu$Jy. The median redshift and depth of Band 6 data are consistent with the findings of \citet{Bethermin+2015}. }
\label{fig:red_distrib}
\end{figure}

We show the flux distribution of our catalog in Figure~\ref{fig:flux_dist}. 57$\%$ of the sources have fluxes fainter than 1~mJy. The binomial uncertainty\footnote{ The binomial uncertainty is defined as  $\sigma_{n_i} = \sqrt{n_i(1-n_i/N)}$ \label{eqn:berr_formula} where $n_i$ is the number of galaxies in bin $i$ and $N$ is the total number of galaxies.} for each flux density bin is shown as the red line at the center of each bin. 
We plot the redshift distribution of the entire sample in Figure~\ref{fig:red_distrib}. Two separate histograms for different frequency bands (blue for Band 6 and red for Band 7) are shown. The median redshifts for bands 6 and 7 are $\langle z \rangle = 2.72$ and 2.57, respectively. 
We also compare the median redshift values from our work with other ALMA surveys. Our median redshift falls within the range of redshifts from other recent studies of SMGs.

\citet{Bethermin+2015} showed that the median redshift  of the sample of dusty galaxies depends significantly on the depth and the observing wavelength of the infrared surveys. They concluded that the  median redshift increases with increasing wavelength up to 2$\,$mm due to the negative $k$-correction. However, increasing the observing depth results in the detections of the lower-redshift, fainter sources. Hence, the observed variation in median values could just be an outcome of the varying depth of the surveys. The 1$\sigma$ rms level in  the $\sim 1.1\,$mm surveys by \citet{Aravena+2016} and  \citet{Dunlop+2017} are 13$\,\mu$Jy~beam$^{-1}$ and 30$\,\mu$Jy~beam$^{-1}$, respectively. The median redshift values of their samples are also lower compared to the other studies ($\langle z \rangle$ = 1.6 and 2.15). The survey by \citet{Brisbin+2017} has a  shallower depth of 150$\,\mu$Jy and, therefore, a higher median redshift of 2.48. Continuing the trend, \citet{Franco+2018} found a population at a median value of $\langle z \rangle$ = 2.9 based on their survey with a sensitivity of  450$\,\mu$Jy. 
In case of our work, the average depths for the Band 6 and 7 are 110$\,\mu$Jy and 300$\,\mu$Jy, respectively. As we have only five galaxies in Band 7, we exclude those galaxies from any further analysis involving the median redshift. Even though we find a slightly larger median redshift for a depth shallower than \citet{Brisbin+2017}, the redshifts and depths are comparable within the margin of errors. Therefore, our redshift distribution is consistent with other studies discussed in this section and the predictions by \citet{Bethermin+2015}.


\subsection{Previously Identified Galaxies}

As mentioned before, 11 ALMA detections in our catalog are not new. If these targets are DFSGs and unlensed, we have still included them in our catalog.
With the availability of multi-wavelength observations from SERVS and our robust 13-band photometry, we can better understand the nature of the previously identified galaxies and the overall population in general. We provide the references to the previous ALMA detection in Column 11 of Table~\ref{tab:almaobs}. Four out of 11 galaxies were serendipitously detected in previous ALMA archival mining studies (\textit{J0216-0506, J0217-0454}: \citealt{fujimoto+2016,Hatsukade+2015}; \textit{J0217-0442}: \citealt{fujimoto+2016};  \textit{J0217-0508}: \citealt{Ono+2014}).

Six of the 11 known galaxies are detected in the ALMA follow-up programs of the bright DFSGs. However, we still include targets in our analysis both the faint and bright SMGs belong to the same category of dusty galaxies. Also, it would be interesting to observe differences in the multi-wavelength properties of these two classes. Three of the six bright SMGs belong to a single Herschel source (\textit{J0219-0524.63, J0219-0524.77, J0219-0524.84}; \citealt{Bussmann+2015}). The Herschel source is resolved into multiple components at the ALMA resolution. The alignment and similar mid-IR colors of these three galaxies indicate an overdense region. The ALMA flux densities of these targets are larger than the rest of our sample galaxies, but we will still include them to compare the physical properties. Rest of the three galaxies (\textit{J0217-0508, J0217-0445, J0217-0504}) are selected from the AzTEC survey. These dusty, $z >3$ bright SMGs are thought to be the progenitors of low redshift massive ellipticals \citep{Ikarashi+2015}. They typically host a compact starburst in their centers, and their luminosities are comparable to nearby ULIRGs.

The remaining galaxy \textit{J0226-0452} from the WMH5 field is a $z \sim 6$ UV-luminous Lyman-break galaxy \citep{Willott+2013}. As some of the sources in our sample are part of the original ALMA program, that could lead to potential environmental biases in our results. However, 10 out of 11 galaxies are either faint SMGs or selected from bright SMG surveys. Only one galaxy from our sample is the primary target of the original ALMA program, and no other galaxy has the same redshift as the primary target (where the redshift of the primary target is known). Given the small amount of overlap between the ALMA primary targets and our faint SMGs, we do not believe that our sample is significantly affected by environmental biases.

Spectroscopic redshifts are available for three targets (\textit{J0226-0452, J0216-0506, J0217-0454}) taken from different studies, and the references are given in the Table~\ref{tab:almaobs}. 
Target \textit{J0216-0506} is detected in most of the studies mentioned above, and \citet{Seko+2016a} have discussed the spectroscopic redshift and ISM properties of that galaxy. It is a member of the sample of $z \sim 1.4$ star-forming main sequence galaxies. CO emission studies with ALMA have indicated larger molecular gas fractions and gas-to-dust ratios than local galaxies. \citet{Banerji+2011} have identified the target \textit{J0217-0454} as Submillimeter Faint Radio Galaxy (SFRG) which are similar to bright SMGs but have hotter dust temperatures. Its spectroscopic redshift ($z = 1.456$) was measured using [OII] 3727 emission line. In all the cases, the spectroscopic redshifts are well within the 95\% confidence interval of our photometric redshifts.

\begin{figure}[htp]
\label{fig:arx_red}
\includegraphics[width=\linewidth]{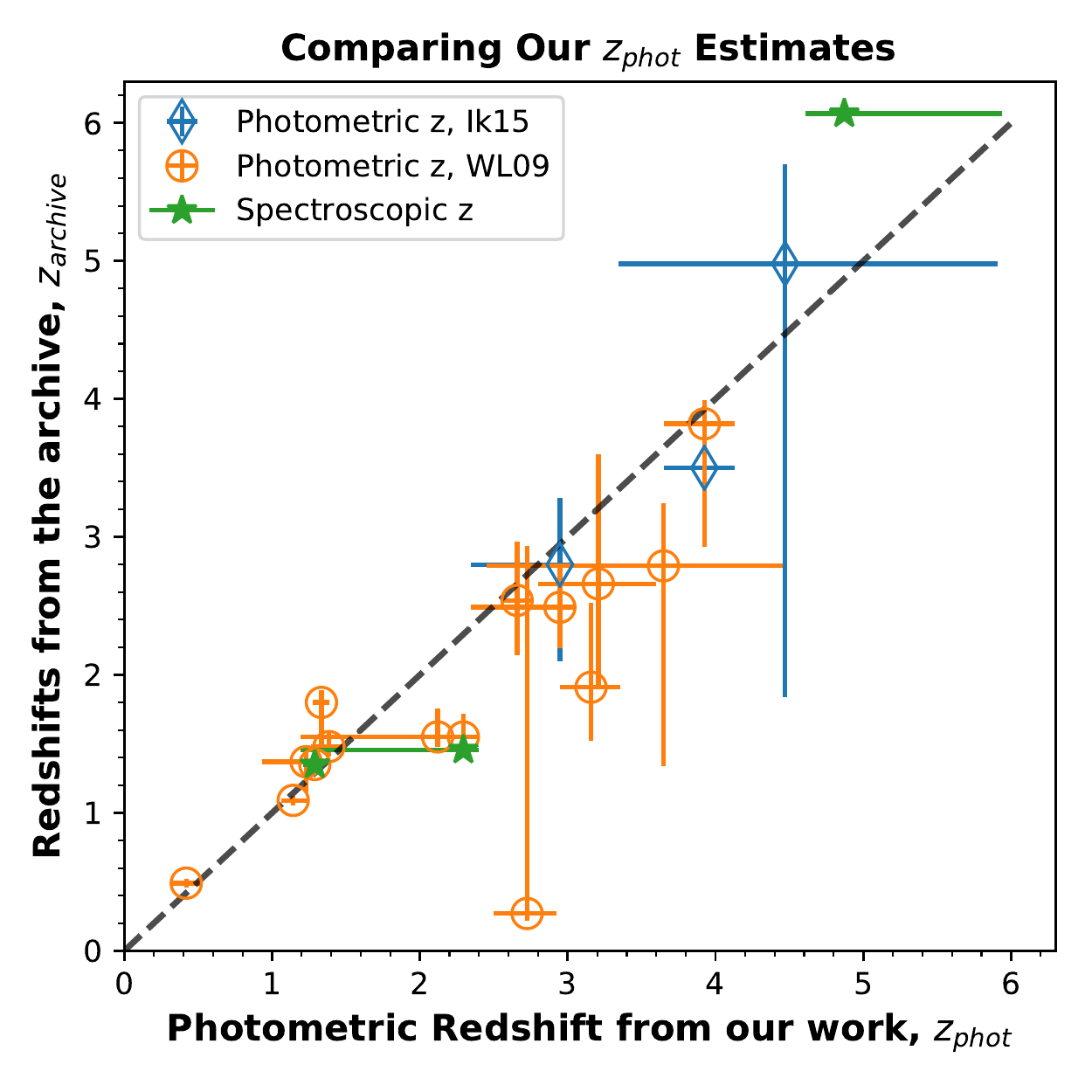}
\caption{A comparison of the photometric redshifts evaluated in this work with redshift estimates from previous studies. A total of 18/26 sources have either photometric or spectroscopic redshifts available in the literature. The green points correspond to sources with spectroscopic redshifts.  The orange and blue points have photometric redshift estimates published by \citet{Williams+2009} and \citet{Ikarashi+2015}, respectively.  The uncertainties from our work are smaller than in previous studies, highlighting our improved photometry through the use of the {\it Tractor} image modeling software. }
\label{fig:arx_comp}
\end{figure}

Figure~\ref{fig:arx_comp} compares the performance of our redshift estimation with archival results. The purpose of this comparison is to conduct a consistency check on the forced photometry technique used in this work. The overall uncertainties on the photo-$z$ estimates are small in most cases. The values are in very good agreement at lower redshifts, $z <2.5$. However, at higher redshifts, we see an increased scatter in the redshift agreement.  This could potentially be due to large uncertainties in the photometric measurements, differences in the filters used for the photo-$z$ estimation, and inherent limitations of the SED templates. 
Our photometry includes $u$-band data which is lacking in the previous studies. Also,  different methods of source deblending in the {\em Spitzer} bands could lead to differences in the photometry. This issue can be resolved using spectroscopic data.

\begin{deluxetable*}{ccccccccccc}[h!]
\tablecaption{ALMA Source Catalog \label{tab:almaobs}}
\tabletypesize\scriptsize
\tablewidth{-2pt}
\tablehead{
\colhead{Source} &  \colhead{IAU ID}  & \colhead{Field} & \colhead{RA} & \colhead{Dec.}  &  \colhead{$\lambda_{obs}$ } & \colhead{$S_{\lambda_{obs}}$} & \colhead{$z_{\rm phot}$} & \colhead{$z_{\rm archival}$} & \colhead{Ref.}  &\colhead{ALMA Ref.}\\
\colhead{} & \colhead{} & \colhead{} & \colhead{(J2000)} & \colhead{(J2000)} & \colhead{(mm)} & \colhead{(mJy)} & \colhead{}  & \colhead{}  & \colhead{} & \colhead{} \\
\colhead{(1)} & \colhead{(2)} & \colhead{(3)} & \colhead{(4)} & \colhead{(5)} & \colhead{(6)} & \colhead{(7)} & \colhead{(8)}  & \colhead{(9)}  & \colhead{(10)} &
\colhead{(11)}
}
\startdata
191313 & J0226-0452 & WMH5 & 02h26m27.00s & -04d52m38.38s & 1.14 & 0.18 $\pm$ 0.04 &  $4.87_{- 0.26}^{+1.07}$ & 6.068$^{\dagger}$ & 4 & a\\
309436 & J0225-0417 & XMMF6 & 02h25m48.03s & -04d17m48.96s & 0.87 & 1.33 $\pm$ 0.53 &  $2.77_{- 0.25}^{+0.14}$ &  \nodata & \nodata & \nodata\\
315617 & J0228-0416 & CLM1 & 02h28m02.58s & -04d16m06.99s & 1.16 & 0.17 $\pm$ 0.03 &  $0.29_{- 0.09}^{+0.26}$ &  \nodata & \nodata & \nodata\\
477360 & J0219-0524.63 & XMMF16 & 02h19m42.63s & -05d24m37.07s & 0.87 & 9.66 $\pm$ 1.22 &  $1.75_{- 0.11}^{+0.16}$ &  2.048 & 1 & b\\
814024 & J0219-0524.52 & XMMF16 & 02h19m42.52s & -05d24m41.27s & 0.87 & 2.61 $\pm$ 0.50 &  $2.49_{- 0.42}^{+0.46}$ &  \nodata & \nodata &\nodata\\
842544 & J0219-0524.77 & XMMF16 & 02h19m42.77s & -05d24m36.43s & 0.87 & 2.81 $\pm$ 0.50 &  $2.57_{- 0.24}^{+0.22}$ &  1.489 & 1 & b\\
911805 & J0219-0524.84 & XMMF16 & 02h19m42.84s & -05d24m35.11s & 0.87 & 2.69 $\pm$ 0.49 &  $2.97_{- 0.12}^{+0.12}$ & 1.489 & 1 & b\\
932297 & J0217-0520 & SXDF1100.277 & 02h17m23.50s & -05d20m08.62s & 1.13 & 2.28 $\pm$ 0.41 & $3.62_{- 0.12}^{+0.12}$ & \nodata & \nodata & \nodata \\
935442 & J0218-0519 & SXDF.220GHZ & 02h18m56.10s & -05d19m51.00s & 1.33 & 0.10 $\pm$ 0.03 &  $1.38_{- 0.12}^{+0.09}$ &  $1.48_{- 0.07}^{+0.10}$ &  1,2 & \nodata \\
971686 & J0217-0511 & UDS16 & 02h17m25.72s & -05d11m03.17s & 1.24 & 0.10 $\pm$ 0.03 & $2.73_{- 0.23}^{+0.20}$ &  $0.27_{- 0.05}^{+2.66}$ & 2 & \nodata \\
978351 & J0217-0510 & UDS16 & 02h17m26.10s & -05d10m58.20s & 1.24 & 0.95 $\pm$ 0.12 &  $2.12_{- 0.34}^{+0.14}$ & $1.55_{- 0.07}^{+0.20}$ &  1,2 & \nodata \\
987090 & J0217-0508 & SXDF1100.027 & 02h17m20.95s & -05d08m37.17s & 1.13 & 1.39 $\pm$ 0.18 &  $2.95_{- 0.60}^{+0.10}$ & $2.80_{- 0.70}^{+0.48}$ &  1,5 & c\\
993676 & J0216-0506 & SXDS5.28019 & 02h16m08.51s & -05d06m15.89s & 1.27 & 0.18 $\pm$ 0.10 &  $1.29_{- 0.08}^{+0.05}$ & 1.348$^{\dagger}$ & 1,2,6 & d,e\\
998253 & J0217-0508 & HIMIKO & 02h17m58.28s & -05d08m30.64s & 1.16 & 0.59 $\pm$ 0.04 &  $1.14_{- 0.08}^{+0.11}$ & $1.09_{- 0.04}^{+0.03}$ &  1,2  &f\\
1002990 & J0216-0503 & SXDF1100.013 & 02h16m47.10s & -05d03m44.54s & 1.13 & 1.46 $\pm$ 0.18 &  $3.03_{- 2.27}^{+0.23}$ & \nodata & \nodata & \nodata\\
1017745 & J0218-0501 & SXDF1100.039 & 02h18m30.98s & -05d01m23.22s & 1.13 & 0.49 $\pm$ 0.17  & $0.42_{- 0.11}^{+0.10}$ &  $0.49_{- 0.03}^{+0.03}$ &  1,2 & \nodata\\\
1048801 & J0217-0454 & SXDS1.59863 & 02h17m46.28s & -04d54m39.77s & 1.23 & 0.69 $\pm$ 0.20  & $2.29_{- 1.10}^{+0.10}$ &  1.456$^{\dagger}$ & 1,2,3 &d,e\\
1094072 & J0217-0445 & SXDF1100.230 & 02h17m59.38s & -04d45m53.13s & 1.13 & 1.54 $\pm$ 0.14 & $3.92_{- 0.27}^{+0.20}$ &  $3.50_{- 0.18}^{+0.40}$ & 1,2,5 & c\\
1106738 & J0217-0442  & SXDS2.22198 & 02h17m53.17s & -04d42m39.61s & 1.27 & 0.44 $\pm$ 0.10  & $0.03_{- 0.02}^{+0.04}$ & \nodata & \nodata & d\\
1266981 & J0217-0520 & SXDF1100.277 & 02h17m23.95s & -05d20m28.02s & 1.13 & 0.55 $\pm$ 0.14  & $2.66_{- 0.10}^{+0.11}$ & $2.54_{- 0.40}^{+0.43}$ &  1 & \nodata\\\
1302443 & J0218-0501 & SXDF1100.039 & 02h18m30.23s & -05d01m21.19s & 1.13 & 0.84 $\pm$ 0.17  & $3.21_{- 0.41}^{+0.39}$ &  $2.66_{- 0.73}^{+0.94}$ &  1 & \nodata\\\
1302615 & J0217-0520 & SXDF1100.250 & 02h17m52.00s & -05d20m32.34s & 1.13 & 0.97 $\pm$ 0.14  & $3.65_{- 1.20}^{+0.81}$ &  $2.79_{- 1.45}^{+0.45}$ &  1 & \nodata\\\
1303410 & J0218-0508 & SXDF1100.109 & 02h18m23.99s & -05d08m11.40s & 1.13 & 0.79 $\pm$ 0.31  & $3.16_{- 0.21}^{+0.20}$ &  $1.91_{- 0.39}^{+0.61}$ &  1 & \nodata\\\
1304155 & J0217-0452 & SXDF1100.063 & 02h17m35.50s & -04d52m11.79s & 1.13 & 0.62 $\pm$ 0.17  & $2.59_{- 0.45}^{+0.44}$ & \nodata & \nodata & \nodata\\\
1307256 & J0218-0457 & SXDF1100.179 & 02h18m43.41s & -04d57m33.05s & 1.13 & 0.79 $\pm$ 0.14  & $5.40_{- 5.38}^{+0.56}$ & \nodata & \nodata & \nodata\ \\
1312163 & J0217-0504 & SXDF1100.110 & 02h17m43.58s & -05d04m10.31s & 1.13 & 2.61 $\pm$ 0.30  & $4.47_{- 1.13}^{+1.44}$ &  $4.98_{- 3.14}^{+0.72}$ &  5 & c\\
\enddata
\tablecomments{Column 1:  Source name.  Column 2:  Source name based on the IAU convention.  Column 3:  Field name indicated in the ALMA archive.  Columns 4-5: Source right ascension and declination. The position corresponds to the peak of the Gaussian source fitted using PyBDSF.  Column 6:  ALMA observing wavelength.  Column 7:  ALMA integrated flux.  Column 8:  Photometric redshift estimated using our \textit{Tractor} forced photometry (Table~\ref{tab:tractor}).  Column 9: Previously published source redshift.  Values marked by the $\dagger$ symbol are spectroscopic redshifts; the rest are photometric redshifts. Column 10: References for literature redshifts and previous detections at all wavelengths: (1) \citet{Robinson+2013}; (2)  \citet{Williams+2009}; (3) \citet{Banerji+2011}; (4) \citet{Willott+2013}; (5) \citet{Ikarashi+2015}; (6) \citet{Seko+2016a}. 
\\
Column 11: References for previous ALMA detections: (a) \citet{Willott+2015}; (b) \citet{Bussmann+2015}; (c) \citet{Ikarashi+2015}; (d) \citet{fujimoto+2016}; (e) \citet{Hatsukade+2013}; (f) \citet{Ono+2014}
}

$\dagger$ Spectroscopic Redshifts
\end{deluxetable*}

\begin{deluxetable*}{cccccccccccccccc}
\rotate
\tabletypesize=\scriptsize
\tablewidth{0pt}
\tablecaption{Multi-band Tractor Photometry \label{tab:tractor}}
\tablehead{
\colhead{Source} &  \colhead{$F_{Ks}$}  & \colhead{$F_{H}$} & \colhead{$F_{J}$} & \colhead{$F_{Y}$}  &  \colhead{$F_{Z}$} & \colhead{$F_{u^{\prime}}$} & \colhead{$F_{g}$} & \colhead{$F_{r}$} & \colhead{$F_{i}$} & \colhead{$F_{z}$} & \colhead{$F_{y}$} & \colhead{$F_{3.6\mu\mathrm{m}}$} & \colhead{$F_{4.5\mu\mathrm{m}}$}\\
\colhead{(1)} & \colhead{(2)} & \colhead{(3)} & \colhead{(4)} & \colhead{(5)} & \colhead{(6)} & \colhead{(7)} & \colhead{(8)}  & \colhead{(9)}  & \colhead{(10)} & \colhead{(11)} & \colhead{(12)} & \colhead{(13)} & \colhead{(14)}
}
\startdata
191313$^{\dagger}$   &    0.37 $\pm$ 0.06 &    0.62 $\pm$ 0.14 &    0.35 $\pm$ 0.03 &    0.47 $\pm$ 0.02 &    0.28 $\pm$ 0.01 &   -0.03 $\pm$ 0.01 &   -0.04 $\pm$ 0.01 &   -0.07 $\pm$ 0.01 &   -0.06 $\pm$ 0.02 &    0.54 $\pm$ 0.04 &    0.38 $\pm$ 0.08 &    1.35 $\pm$ 0.07 &    0.81 $\pm$ 0.10 \\ 
309436$^{\dagger}$   &    7.43 $\pm$ 0.83 &    6.48 $\pm$ 0.22 &    2.39 $\pm$ 0.10 &    1.38 $\pm$ 0.07 &    1.05 $\pm$ 0.05 &    0.04 $\pm$ 0.02 &    0.28 $\pm$ 0.02 &    0.90 $\pm$ 0.03 &    0.84 $\pm$ 0.05 &    1.06 $\pm$ 0.08 &    1.32 $\pm$ 0.14 &    7.00 $\pm$ 0.14 &   12.92 $\pm$ 0.13 \\
315617$^{\dagger}$   &    0.51 $\pm$ 0.07 &    0.91 $\pm$ 0.16 &    0.81 $\pm$ 0.04 &    0.78 $\pm$ 0.03 &    0.55 $\pm$ 0.02 &    \nodata         &    0.01 $\pm$ 0.01 &    0.20 $\pm$ 0.01 &    0.31 $\pm$ 0.02 &    0.42 $\pm$ 0.04 &    0.54 $\pm$ 0.07 &   -0.20 $\pm$ 0.07 &   -0.92 $\pm$ 0.10 \\
477360   &    6.83 $\pm$ 0.72 &    3.05 $\pm$ 0.12 &    1.23 $\pm$ 0.07 &    0.23 $\pm$ 0.04 &   -0.31 $\pm$ 0.02 &   -0.01 $\pm$ 0.08 &   -0.10 $\pm$ 0.01 &   -0.21 $\pm$ 0.02 &   -0.02 $\pm$ 0.02 &   -0.01 $\pm$ 0.04 &    0.36 $\pm$ 0.08 &   13.61 $\pm$ 0.06 &   24.10 $\pm$ 0.09 \\ 
814024   &    1.77 $\pm$ 0.27 &    0.93 $\pm$ 0.06 &    0.21 $\pm$ 0.04 &    0.05 $\pm$ 0.02 &    0.02 $\pm$ 0.01 &   -0.03 $\pm$ 0.04 &    0.01 $\pm$ 0.01 &    0.03 $\pm$ 0.01 &    0.01 $\pm$ 0.01 &    0.02 $\pm$ 0.02 &    0.10 $\pm$ 0.04 &    5.37 $\pm$ 0.08 &    7.12 $\pm$ 0.11 \\ 
842544   &    2.20 $\pm$ 0.29 &    1.68 $\pm$ 0.07 &    0.31 $\pm$ 0.03 &    0.18 $\pm$ 0.02 &    0.12 $\pm$ 0.01 &    0.01 $\pm$ 0.04 &   -0.01 $\pm$ 0.01 &    0.09 $\pm$ 0.01 &    0.05 $\pm$ 0.01 &    0.05 $\pm$ 0.02 &   -0.00 $\pm$ 0.04 &   10.84 $\pm$ 0.07 &   15.77 $\pm$ 0.09 \\ 
911805   &   11.12 $\pm$ 1.12 &    9.23 $\pm$ 0.30 &    3.89 $\pm$ 0.14 &    3.61 $\pm$ 0.12 &    3.04 $\pm$ 0.10 &    0.34 $\pm$ 0.10 &    1.40 $\pm$ 0.04 &    1.95 $\pm$ 0.06 &    2.54 $\pm$ 0.08 &    2.94 $\pm$ 0.10 &    3.04 $\pm$ 0.14 &   24.17 $\pm$ 0.07 &   32.00 $\pm$ 0.09 \\ 
932297   &    0.71 $\pm$ 0.09 &    0.59 $\pm$ 0.06 &    0.39 $\pm$ 0.05 &    0.51 $\pm$ 0.09 &    0.47 $\pm$ 0.02 &    0.00 $\pm$ 0.04 &    0.15 $\pm$ 0.01 &    0.46 $\pm$ 0.02 &    0.44 $\pm$ 0.02 &    0.52 $\pm$ 0.03 &    0.49 $\pm$ 0.04 &   -0.14 $\pm$ 0.07 &    0.17 $\pm$ 0.08 \\ 
935442   &   16.62 $\pm$ 1.54 &   10.80 $\pm$ 0.33 &    5.54 $\pm$ 0.18 &    4.02 $\pm$ 0.12 &    1.99 $\pm$ 0.06 &    0.11 $\pm$ 0.06 &    0.31 $\pm$ 0.01 &    0.57 $\pm$ 0.02 &    1.06 $\pm$ 0.03 &    1.83 $\pm$ 0.06 &    3.06 $\pm$ 0.11 &   26.82 $\pm$ 0.04 &   29.01 $\pm$ 0.05 \\ 
971686   &    0.44 $\pm$ 0.07 &    0.64 $\pm$ 0.05 &    0.35 $\pm$ 0.04 &    0.42 $\pm$ 0.07 &    0.29 $\pm$ 0.01 &    0.05 $\pm$ 0.04 &    0.30 $\pm$ 0.01 &    0.37 $\pm$ 0.01 &    0.31 $\pm$ 0.01 &    0.30 $\pm$ 0.02 &    0.27 $\pm$ 0.03 &    0.61 $\pm$ 0.03 &    1.07 $\pm$ 0.05 \\ 
978351   &   19.63 $\pm$ 1.82 &   10.49 $\pm$ 0.33 &    4.96 $\pm$ 0.17 &    2.16 $\pm$ 0.08 &    1.16 $\pm$ 0.04 &    0.11 $\pm$ 0.09 &    0.18 $\pm$ 0.01 &    0.44 $\pm$ 0.02 &    0.77 $\pm$ 0.03 &    1.17 $\pm$ 0.05 &    1.86 $\pm$ 0.09 &   39.81 $\pm$ 0.04 &   48.34 $\pm$ 0.06 \\ 
987090   &    7.04 $\pm$ 0.67 &    3.63 $\pm$ 0.12 &    0.73 $\pm$ 0.05 &    0.29 $\pm$ 0.02 &    0.18 $\pm$ 0.01 &    0.01 $\pm$ 0.05 &   -0.01 $\pm$ 0.00 &    0.01 $\pm$ 0.01 &    0.05 $\pm$ 0.01 &    0.13 $\pm$ 0.02 &    0.22 $\pm$ 0.04 &   15.35 $\pm$ 0.04 &   14.16 $\pm$ 0.06 \\ 
993676   &   14.72 $\pm$ 1.35 &   11.32 $\pm$ 0.35 &    6.05 $\pm$ 0.19 &    5.67 $\pm$ 0.17 &    3.98 $\pm$ 0.12 &    0.48 $\pm$ 0.06 &    0.83 $\pm$ 0.03 &    1.36 $\pm$ 0.04 &    2.06 $\pm$ 0.06 &    3.79 $\pm$ 0.12 &    4.79 $\pm$ 0.15 &   19.85 $\pm$ 0.06 &   21.56 $\pm$ 0.09 \\ 
998253   &   20.95 $\pm$ 1.93 &   12.57 $\pm$ 0.39 &    6.75 $\pm$ 0.23 &    5.56 $\pm$ 0.17 &    4.35 $\pm$ 0.14 &    0.48 $\pm$ 0.10 &    0.82 $\pm$ 0.03 &    1.28 $\pm$ 0.04 &    2.21 $\pm$ 0.07 &    4.30 $\pm$ 0.13 &    5.09 $\pm$ 0.17 &   37.95 $\pm$ 0.08 &   35.73 $\pm$ 0.08 \\ 
1002990  &   -0.06 $\pm$ 0.06 &    0.19 $\pm$ 0.04 &    0.21 $\pm$ 0.03 &    0.01 $\pm$ 0.02 &    0.13 $\pm$ 0.03 &    0.02 $\pm$ 0.03 &    0.06 $\pm$ 0.00 &    0.08 $\pm$ 0.01 &    0.10 $\pm$ 0.01 &    0.15 $\pm$ 0.01 &    0.16 $\pm$ 0.03 &    0.53 $\pm$ 0.06 &    0.32 $\pm$ 0.09 \\ 
1017745  &   19.10 $\pm$ 1.77 &   16.91 $\pm$ 0.52 &   11.27 $\pm$ 0.35 &   11.30 $\pm$ 0.34 &    9.29 $\pm$ 0.28 &    1.04 $\pm$ 0.10 &    2.00 $\pm$ 0.06 &    4.96 $\pm$ 0.15 &    7.11 $\pm$ 0.21 &    9.10 $\pm$ 0.28 &   10.68 $\pm$ 0.33 &    9.72 $\pm$ 0.12 &    9.64 $\pm$ 0.12 \\ 
1048801  &   13.33 $\pm$ 1.27 &    7.36 $\pm$ 0.24 &    3.83 $\pm$ 0.14 &    2.19 $\pm$ 0.08 &    1.56 $\pm$ 0.05 &    0.19 $\pm$ 0.08 &    0.47 $\pm$ 0.02 &    0.68 $\pm$ 0.02 &    1.06 $\pm$ 0.04 &    1.62 $\pm$ 0.06 &    1.88 $\pm$ 0.08 &   26.22 $\pm$ 0.04 &   32.26 $\pm$ 0.07 \\ 
1094072  &    1.42 $\pm$ 0.24 &    0.51 $\pm$ 0.05 &    0.22 $\pm$ 0.04 &    0.23 $\pm$ 0.02 &    0.21 $\pm$ 0.01 &    0.02 $\pm$ 0.04 &    0.02 $\pm$ 0.00 &    0.12 $\pm$ 0.01 &    0.12 $\pm$ 0.01 &    0.16 $\pm$ 0.02 &    0.17 $\pm$ 0.03 &    3.13 $\pm$ 0.03 &    3.35 $\pm$ 0.06 \\ 
1106738  &    0.75 $\pm$ 0.11 &    1.19 $\pm$ 0.25 &    0.86 $\pm$ 0.06 &    0.99 $\pm$ 0.04 &    0.94 $\pm$ 0.03 &    0.24 $\pm$ 0.06 &    0.68 $\pm$ 0.02 &    0.91 $\pm$ 0.03 &    0.96 $\pm$ 0.03 &    1.01 $\pm$ 0.04 &    1.00 $\pm$ 0.07 &    0.00 $\pm$ 0.05 &   -0.08 $\pm$ 0.06 \\ 
1266981  &   10.22 $\pm$ 1.02 &    5.58 $\pm$ 0.19 &    0.74 $\pm$ 0.09 &    0.56 $\pm$ 0.04 &    0.23 $\pm$ 0.02 &    0.02 $\pm$ 0.07 &    0.00 $\pm$ 0.01 &    0.07 $\pm$ 0.01 &    0.14 $\pm$ 0.02 &    0.20 $\pm$ 0.03 &    0.41 $\pm$ 0.06 &   21.25 $\pm$ 0.05 &   27.94 $\pm$ 0.06 \\ 
1302443  &    1.43 $\pm$ 0.27 &    0.57 $\pm$ 0.06 &    0.14 $\pm$ 0.05 &   -0.05 $\pm$ 0.03 &   -0.01 $\pm$ 0.02 &   -0.02 $\pm$ 0.04 &    0.02 $\pm$ 0.01 &    0.04 $\pm$ 0.01 &    0.03 $\pm$ 0.01 &    0.22 $\pm$ 0.03 &   -0.01 $\pm$ 0.04 &    3.96 $\pm$ 0.13 &    6.70 $\pm$ 0.16 \\ 
1302615  &    1.15 $\pm$ 0.27 &    0.30 $\pm$ 0.05 &   -0.04 $\pm$ 0.05 &    0.05 $\pm$ 0.02 &   -0.11 $\pm$ 0.01 &   -0.01 $\pm$ 0.04 &    0.00 $\pm$ 0.01 &    0.02 $\pm$ 0.01 &    0.01 $\pm$ 0.01 &    0.04 $\pm$ 0.02 &    0.06 $\pm$ 0.04 &    3.34 $\pm$ 0.03 &    4.86 $\pm$ 0.05 \\ 
1303410  &    1.87 $\pm$ 0.28 &    0.96 $\pm$ 0.06 &    0.31 $\pm$ 0.04 &    0.02 $\pm$ 0.02 &    0.11 $\pm$ 0.01 &   -0.01 $\pm$ 0.04 &    0.00 $\pm$ 0.00 &   -0.00 $\pm$ 0.01 &    0.01 $\pm$ 0.01 &    0.05 $\pm$ 0.02 &   -0.02 $\pm$ 0.03 &    4.82 $\pm$ 0.04 &    5.93 $\pm$ 0.07 \\ 
1304155  &    1.60 $\pm$ 0.28 &    0.88 $\pm$ 0.07 &    0.16 $\pm$ 0.06 &    0.12 $\pm$ 0.04 &    0.03 $\pm$ 0.04 &   -0.05 $\pm$ 0.06 &    0.01 $\pm$ 0.01 &    0.04 $\pm$ 0.01 &    0.03 $\pm$ 0.01 &    0.01 $\pm$ 0.02 &    0.08 $\pm$ 0.04 &    5.88 $\pm$ 0.07 &    8.44 $\pm$ 0.09 \\ 
1307256  &    0.61 $\pm$ 0.26 &   -0.14 $\pm$ 0.10 &   -0.23 $\pm$ 0.12 &    0.03 $\pm$ 0.06 &    0.07 $\pm$ 0.05 &   -0.02 $\pm$ 0.04 &   -0.01 $\pm$ 0.00 &   -0.02 $\pm$ 0.01 &   -0.01 $\pm$ 0.01 &   -0.03 $\pm$ 0.02 &   -0.04 $\pm$ 0.03 &    0.30 $\pm$ 0.15 &    3.02 $\pm$ 0.11 \\ 
1312163  &    0.86 $\pm$ 0.28 &    0.10 $\pm$ 0.06 &   -0.01 $\pm$ 0.05 &    0.10 $\pm$ 0.03 &    0.04 $\pm$ 0.02 &   -0.05 $\pm$ 0.05 &    0.00 $\pm$ 0.01 &   -0.03 $\pm$ 0.01 &   -0.02 $\pm$ 0.01 &   -0.03 $\pm$ 0.02 &   -0.09 $\pm$ 0.04 &    1.77 $\pm$ 0.12 &    2.99 $\pm$ 0.13 \\
\enddata
\tablecomments{All fluxes and uncertainties are given in units of $\mu$Jy.  Column 1: Source name.  Column 2: $Ks$-band flux from the VIDEO survey.  Column 3: $H$-band flux from the VIDEO survey.  Column 4: $J$-band flux from the VIDEO survey. Column 5: $Y$-band flux from the VIDEO survey. Column 6: $Z$-band flux from the VIDEO survey. Column 7: $u^{\prime}$-band flux from the wide tier of the CFHTLS survey (CFHTLS-W1). Column 8: $g$-band flux from the Ultradeep tier of DR1 of the HSC survey.  Column 9: $r$-band flux from the Ultradeep tier of DR1 of the HSC survey. Column 10: $i$-band flux from the Ultradeep tier of DR1 of the HSC survey. Column 11: $z$-band flux from the Ultradeep tier of DR1 of the HSC survey. Column 12: $y$-band flux from the Ultradeep tier of DR1 of the HSC survey. Column 13: 3.6$\mu$m {\it Spitzer} IRAC flux from the DeepDrill survey. Column 14: 4.5$\mu$m {\it Spitzer} IRAC flux from the DeepDrill survey.
\\
$\dagger$ The CFHTLS flux is based on the deeper CFHTLS-D1 tier and the HSC flux is based on the shallower Deep tier.\\
}
\end{deluxetable*}


\subsection{Optical-NIR Triple Color Selection }

\citet{Chen+OIR} have devised a selection technique to identify bright as well as faint SMGs using a training dataset from the UKIDSS-UDS field. The method is called Optical-Infrared Triple Color selection (OIRTC) and uses three optical-NIR colors:$z-K_{s}$, $K_{s}-[3.6]$,  and $[3.6]-[4.5]$. The color cuts are defined such that the mean SMG fraction redder than the color threshold is at least 5$\%$. The training sample contains  ALMA and 850 $\mu$m SCUBA-2 observations of bright SMGs. By combining radio and optical-NIR selection techniques, their identification is about 83$\%$ complete. \citet{Chen+2016} have utilized this technique to identify faint SMGs with expected fluxes, $S_{850\,\mu m} < 2$ mJy.

\begin{figure*}
    \gridline{\fig{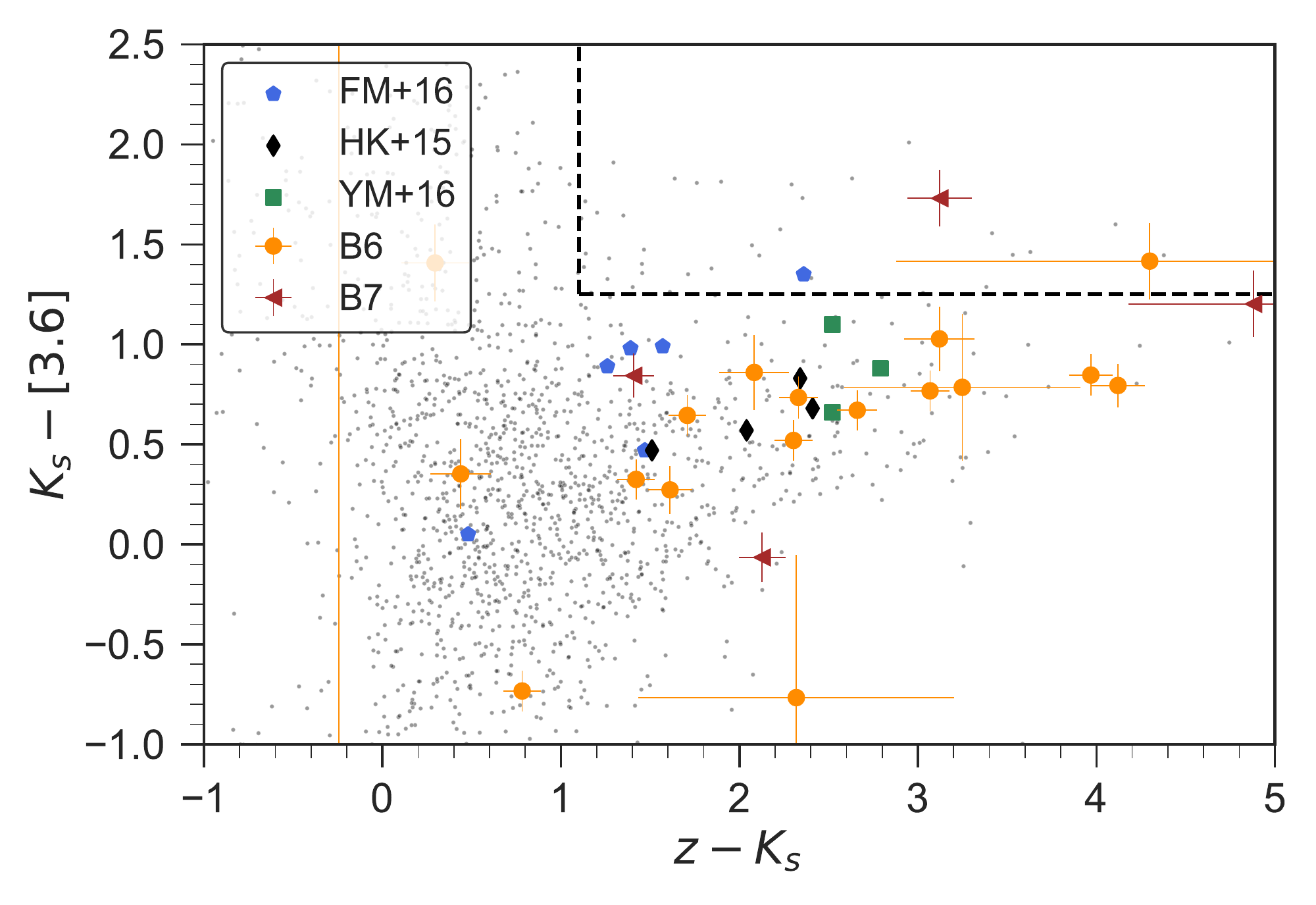}{0.46\linewidth}{(a)}
          \fig{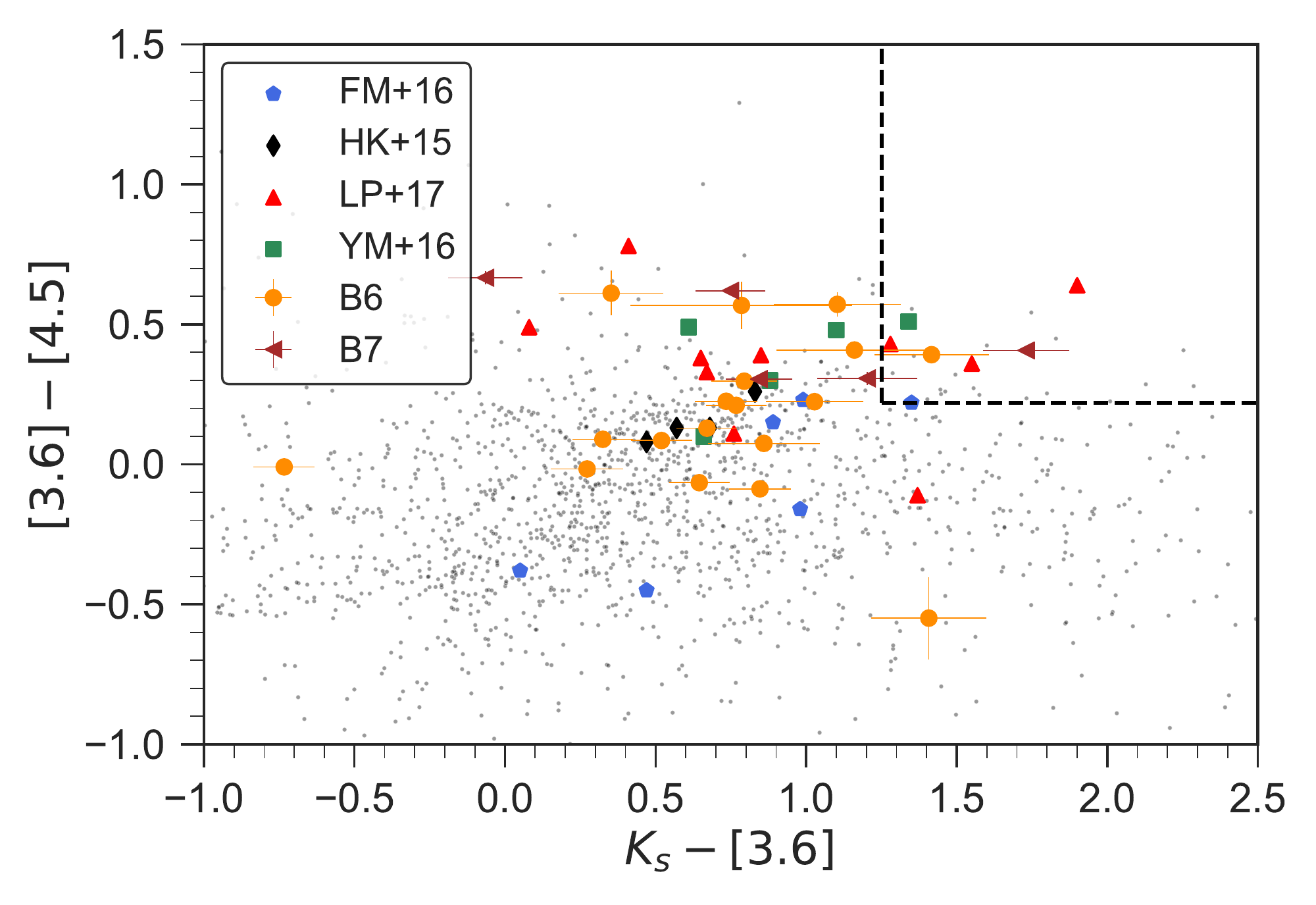}{0.46\linewidth}{(b)}
          }
     \gridline{
          \fig{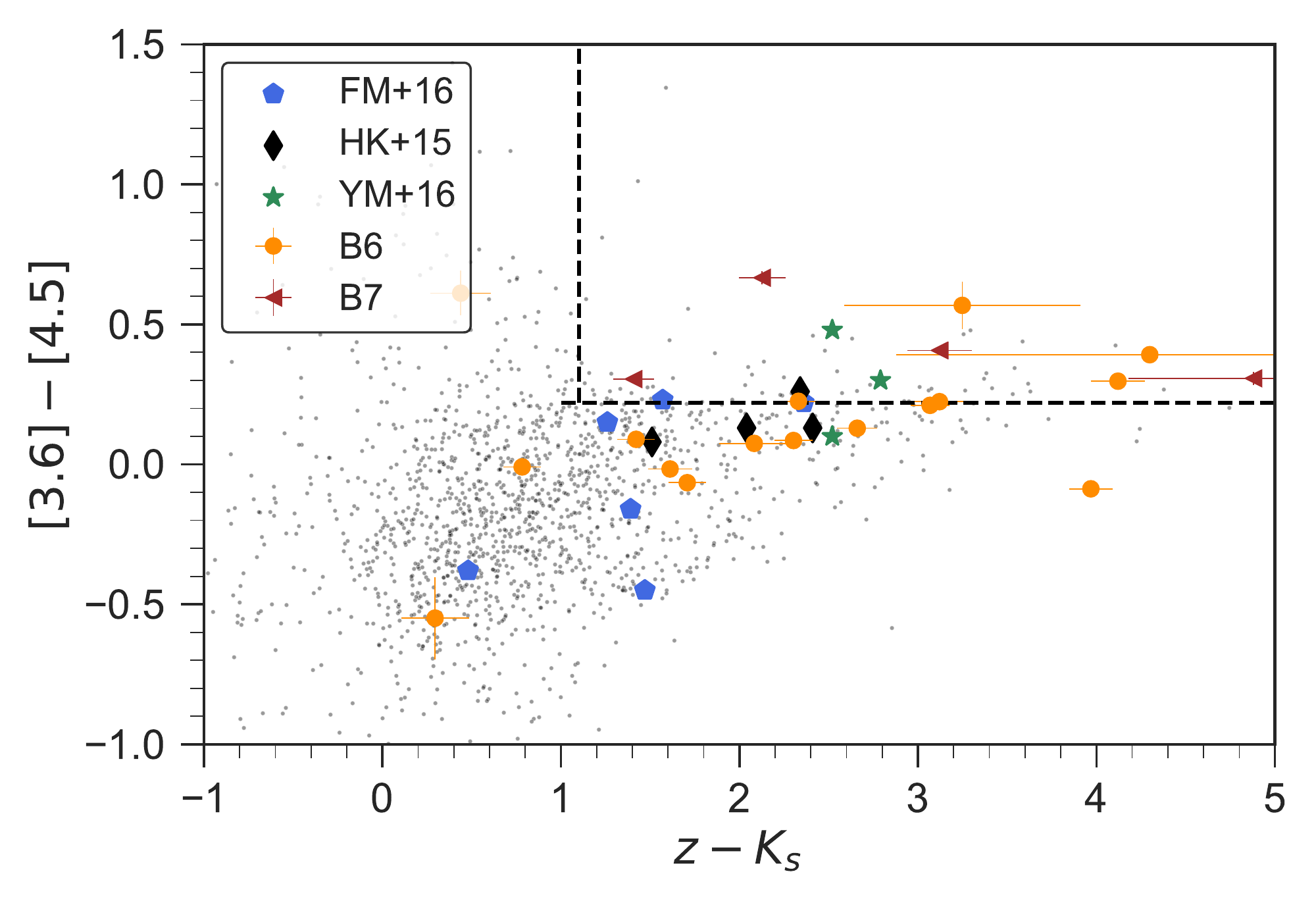}{0.46\linewidth}{(c)}
          }
	\caption{The plot shows the SMG OIRTC color selection criteria defined in \citet{Chen+OIR}. Each sub-figure is a color-color plot of the combination of $z$, $K_s$, $[3.6]$, and $[4.5]$ filters. Bright SMGs tend to occupy the reddest part of the color space due to the presence of dust. The black dotted line in each sub-figure is a color cut-off estimated by \citet{Chen+OIR} using an ALMA sample of bright SMGs. The fraction of SMGs in the population redder than the color threshold is at least 5$\%$. The yellow filled circles (B6), and brown horizontal triangles (B7) represent Band 6 and 7 sources in our sample, respectively. Other faint SMG samples are shown as follows:- YM+16 (Green squares) - \citet{Yamaguchi+16}; HK+15 (Black diamonds) - \citet{Hatsukade+2015}; FM+16 (Blue pentagons) - \citet{fujimoto+2016}; LP+17 (Red vertical triangles) - \citet{Laporte+17}. The light gray symbols  in the background show the field non-SMG galaxies found within all ALMA pointings. 
Our sample occupies a bluer color space than the OIRTC selection cut-off mainly due to the mail$Ks-[3.6]\,\mu m$ color. This shows that the optical-NIR color properties of our faint SMG population are different from the bright SMGs. \\}
	\label{fig:oirtc}
\end{figure*}

Here, we compare the optical-IR colors of our faint SMG sample with the color selection method described in \citet{Chen+OIR}. Figure~\ref{fig:oirtc} illustrates the OIRTC color selection, where different colored symbols represent our sample along with faint SMG samples of \citet{fujimoto+2016}, \citet{Hatsukade+2015}, \citet{Laporte+17}, and \citet{Yamaguchi+16}.
A comparison with other samples will allow us to evaluate the significance of our results and also to check the consistency between different methods adopted in all of the studies. We see that almost our entire sample occupies the color space outside the selection cuts defined by the OIRTC technique.  Furthermore,  the galaxies from the comparison samples also lie mostly outside the OIRTC color selection.  The light-grey symbols in Figure~\ref{fig:oirtc} are the non-SMG field galaxies found within the search radius of all the ALMA pointings used in our study. These field galaxies will allow us to check whether the colors of faint SMGs are systematically redder.

We observe that 75\% of faint SMGs satisfy the $z-K_s$ color criterion of having redder colors than the OIRTC $z-K_s$ color criterion. Whereas only 47\% and 18\% of  the faint SMG population satisfy the OIRTC $ [3.6]\,\mu m - [4.5]\,\mu m$ and $K_s - [3.6]\,\mu m$ color criteria, respectively. \citet{Laporte+17} and \citet{Cowie+18} have also analyzed the OIRTC criteria for their sample sources. Both the studies found that majority of their samples do not satisfy the OIRTC $K_s-[3.6]\,\mu m$ cut. A few factors could be responsible for the differences. The primary extrinsic factor is the variations in the photometry techniques used to obtain colors. \citet{Laporte+17} pointed out that the disagreement of the $K_s-[3.6]$ colors between their and  \citet{Chen+OIR} sample was mainly due to the differences in the \textit{Spitzer} IRAC aperture correction methods. 
When the OIRTC color criteria were re-calibrated for their field galaxies, the new color-cuts were 80\% complete. The \textit{Tractor} forced photometry presented in our work takes the point spread function of each band into account, thus removing the need for aperture corrections. Nevertheless, we still find a significant fraction of the sample occupying bluer Ks-[3.6] colors, which may be due to the differences in the intrinsic properties of the faint and bright SMGs.
Hence, additional checks are needed before directly applying color cuts from the other studies. It would be useful to perform future studies which combine all the different faint SMGs and provide consistent photometry for all the comparison samples.

As mentioned above, the other reason could be the intrinsic differences in the SEDs of bright and faint SMGs. The OIRTC color selection method is trained using bright SMGs, and the faint population in \citet{Chen+2016} does not have confirmed interferometric detections. Hence, the technique is targeting faint galaxies having SEDs similar to the bright SMGs. 
To further support this, \citet{Hatsukade+2015} found bluer colors for their sample of faint SMGs. Therefore, additional study is needed, including spectroscopic follow-up, in order to better understand the colors of faint SMGs.


\subsection{ The UVJ color-color plot \label{sec:uvj}}

At higher redshifts, where it is difficult to classify galaxies based on their morphologies, classifications via rest-frame broad-band colors have proven to be useful. Star-forming and quiescent galaxies show a bi-modality in rest-frame colors up to $z \sim 2.5$ in the UVJ color-color diagram
 \citep{Labbe+2005,Wuyts+2007,Williams+2009,Whitaker+2011,Brammer+2011,Patel+2011}. The quiescent galaxies form a red clump called the "red sequence", and star-forming galaxies fall on a diagonal line. The redder galaxies on the star-forming main sequence have higher dust extinction. The $U-V$ and $V-J$ colors of the DFSGs are both reddened by dust extinction. However, in the case of quiescent galaxies, only the $U-V$ colors are reddened due to the Balmar/4000 \AA~break, and the $V-J$ colors are bluer  as compared to the $U-V$ \citep{Chen+OIR}. Therefore, the $V-J$ color is a good proxy for dust extinction, and the two colors can separate DFSGs and red sequence quiescent galaxies up to $z \sim 2.5$.

Figure \ref{fig:uvj} shows the rest frame UVJ color diagram for our sample. The rest frame colors are estimated using EAZY as explained in Section~\ref{sec:photoz}. To estimate the uncertainties in the U, V, and J  bands, we find the two nearest filters for each band, and add their errors in quadrature. Then, we calculate the uncertainties of the rest-frame colors using a simple error propagation rule. Here, we consider only measurement errors and exclude the errors in the SED templates and fitting. 

The quiescent galaxy color cuts are taken from \citet{Williams+2009}. The color selection is defined as follows:

\begin{eqnarray}
U-V > 1.3  \\
V-J < 1.6 \\
U-V > 0.88 \times (V-J) + 0.49 
\end{eqnarray}
The third criterion has a small dependence on redshift, and we used the equation for the redshift bin $1<z<2$. Five out of 26 galaxies are located within the quiescent box, and the rest of the sample has colors similar to the star-forming galaxies. When we compare the bi-color sequence from \citet{Williams+2009}, we see that our star-forming sample occupies redder $U-V$ and $V-J$ colors indicating the dusty and high-redshift nature of our sample.

\begin{figure}
\includegraphics[width=\linewidth]{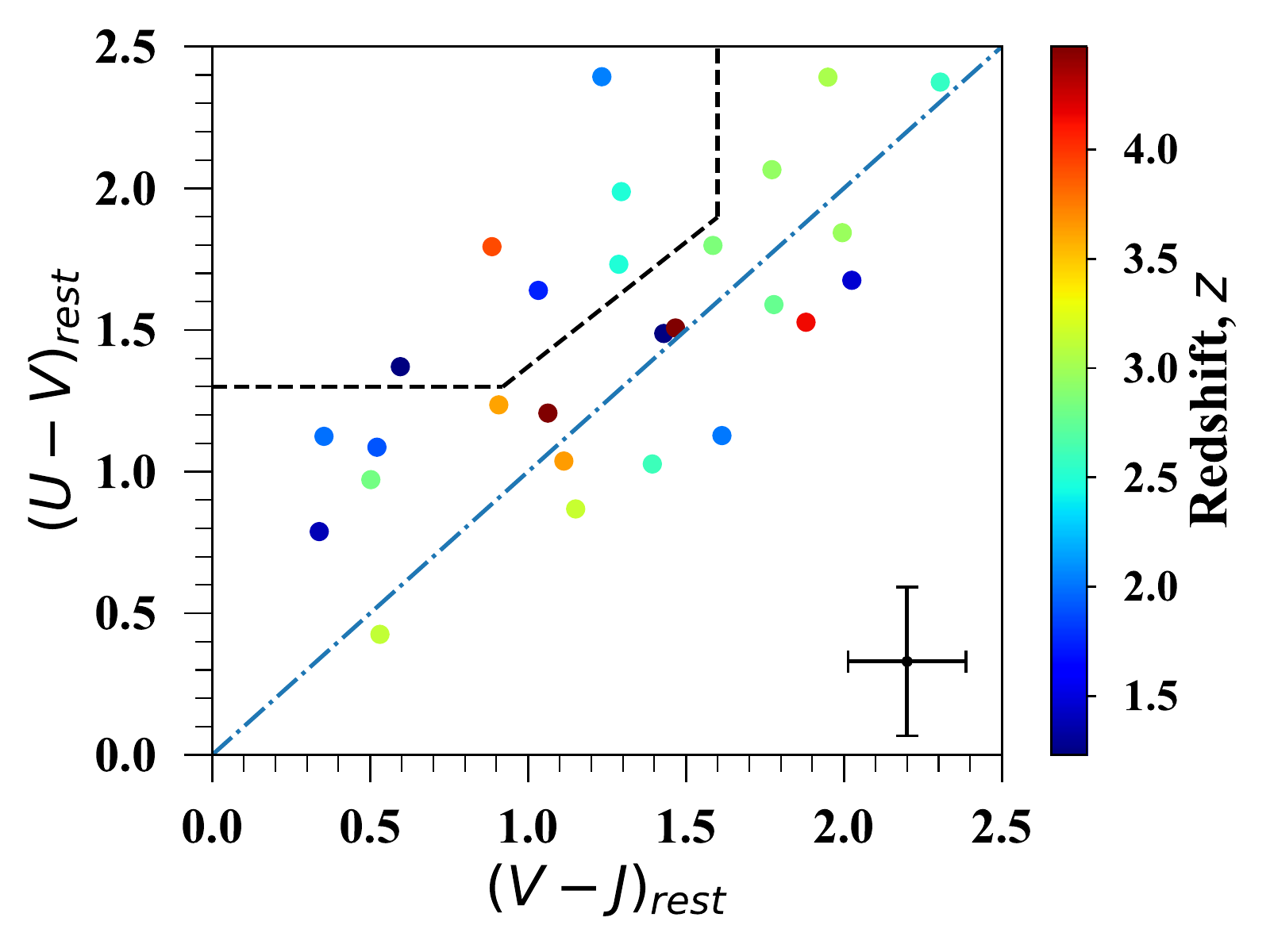}
\caption{UVJ color selection diagram. The symbols are color coded based on their photometric redshifts and the color bar is shown on the right. Five out of the 26 galaxies lie in the quiescent galaxy clump. The blue dotted line is a proportionality line and is plotted to identify the diagonal track occupied by the star-forming galaxies. The black error-bar in the bottom-right corner shows the average uncertainties in the color estimation.}
\label{fig:uvj}
\end{figure}

\subsection{Stacking of Undetected ALMA Sources}

 We used stacking to investigate the nature of the faint population that was not individually detected in our photometric catalog. As the rate of contamination of the ALMA catalog by noise fluctuations is expected to be very small above 5$\sigma$, we stacked the DeepDrill IRAC data at the positions of the $>5\sigma$ sources (60/176). We divided our sample into sources that were bright ($>1\,$mJy), or faint ($<1\,$mJy)
 We only detected $>3 \sigma$ emission in the IRAC 4.5$\,\mu$m channel stack of the faint sources (Table \ref{tab:stacking}) at the level of 0.34$\,\mu$Jy ($AB=25.3$). This indicates that the majority of the undetected population is extremely faint in the IRAC bands ($AB \stackrel{>}{_{\sim}}25$), and is either of low stellar mass ($\stackrel{<}{_{\sim}} 10^{10}\,M_{\odot}$ at $z=2$) and/or highly reddened.

\begin{table}
\caption{Results of stacking $>5 \sigma$ ALMA objects that were undetected in the IRAC bands}\label{tab:stacking}
\begin{tabular}{lccc}
Stack & Number & 3.6$\mu$m & 4.5$\mu$m\\
      &        & AB mag & AB mag \\\hline
All undetected & 60    & $>$26.1 & $>$26.1\\
$>1$mJy        & 44    & $>$25.8 & $>$25.9\\
$<1$mJy        & 16   & $>$25.5 & 25.3\\
\end{tabular}
\end{table}

\section{Summary and Conclusions}
\label{sec:summary}

In this paper, we have successfully demonstrated the use of archival ALMA observations to search for faint SMGs. The ALMA detections in the XMM-LSS field greater than 3.9$\sigma$ were cross-matched with the deep optical-NIR observations taken from SERVS/DeepDrill, VIDEO, CFHTLS, and HSC. We identify 26 sources with NIR counterparts, 15 of which are newly-identified faint SMGs. To further investigate the nature of this cosmologically important population, we have analyzed the basic properties of our sample. Of the total 26 sources, there are 16 faint SMGs ($S_{1.1\,mm} <$ 1~mJy) with a median flux of 0.57~mJy and 10 bright SMGs with a median flux of 2.44~mJy. Robust 13-band forced photometry using the {\it Tractor} image modeling code is available for our entire catalog. 

The resulting photometric measurements were used to estimate photometric redshifts and rest-frame colors. Our sources have redshifts in the range of $0.4 < z_{\mathrm{phot}} < 5.3$, with a median photometric redshifts of $\langle z \rangle$ = 2.72 and 2.57 for Band 6 and 7, respectively. The median redshift and the average depth of our search are in good agreement with the predictions given in \citet{Bethermin+2015}. 

We performed an optical-NIR triple-color selection that showed that most of our sample galaxies have bluer colors than the redder bright SMGs.  Different color properties of faint SMGs could indicate different physical properties compared to their brighter counterparts. Based on the rest-frame UVJ colors, we found that most of the galaxies in our sample form the star-forming diagonal track on the UVJ diagram. Their colors are consistent with star-forming main sequence galaxies. 

ALMA has made the discovery of this population of DSFGs possible. We will continue mining publicly available ALMA archival observations, and expand our search to the remaining five SERVS/DeepDrill fields to find a large sample of faint SMGs. We will include robust optical-NIR photometry along with far-IR \textit{Herschel} observations to perform SED modeling of all the targets and estimate stellar masses, star formation rates, and other physical properties. Pilot studies (including this work) are unveiling the nature of the faint SMGs that dominate the CIB. Future large-scale surveys are essential to understanding the role of these galaxies in shaping galaxy evolution.

\section*{Acknowledgments}

The National Radio Astronomy Observatory is a facility of the National Science Foundation operated under cooperative agreement by Associated Universities, Inc.  Support for this work was provided by the NSF through the Grote Reber Fellowship Program administered by Associated Universities, Inc./National Radio Astronomy Observatory. This paper makes use of the following ALMA data: ADS/JAO.ALMA  \#2011.0.00115.S, 
 \#2011.0.00648.S,  \#2011.0.00539.S,  \#2012.1.00326.S,  \#2012.1.00374.S,  \#2012.1.00934.S,  \#2013.1.00815.S,  \#2013.A.00021.S,  \#2015.1.01105.S. ALMA is a partnership of ESO (representing its member states), NSF (USA) and NINS (Japan), together with NRC (Canada) and NSC and ASIAA (Taiwan) and KASI (Republic of Korea), in cooperation with the Republic of Chile. The Joint ALMA Observatory is operated by ESO, AUI/NRAO and NAOJ.  This work is based on observations made with the {\it Spitzer} Space Telescope, which is operated by the Jet Propulsion Laboratory, California Institute of Technology under a contract with NASA.  K.N. acknowledges support provided by the grant associated with {\it Spitzer} proposal 11086. J.A. gratefully acknowledges support from the Science and Technology Foundation (FCT, Portugal) through the research grants PTDC/FIS-AST/2194/2012,  PTDC/FIS-AST/29245/2017, UID/FIS/04434/2013 and UID/FIS/04434/2019.

The authors have made use of {\sc Astropy}, a community-developed core {\sc Python} package for Astronomy \citet{astropy+13}.  We also used MONTAGE, which is funded by the National Science Foundation under Grant Number ACI-1440620, and was previously funded by the National Aeronautics and Space Administration's Earth Science Technology Office, Computation Technologies Project, under Cooperative Agreement Number NCC5-626 between NASA and the California Institute of Technology. This research has made use of the NASA/IPAC Extragalactic Database (NED) which is operated by the Jet Propulsion Laboratory, California Institute of Technology, under contract with the National Aeronautics and Space Administration.

\vspace{5mm}
\facility{ALMA, \textit{Spitzer}, CFHT, VISTA, Subaru}





\appendix
\begin{figure*}[htpb!]
\centering
\includegraphics[height=0.9\textheight]{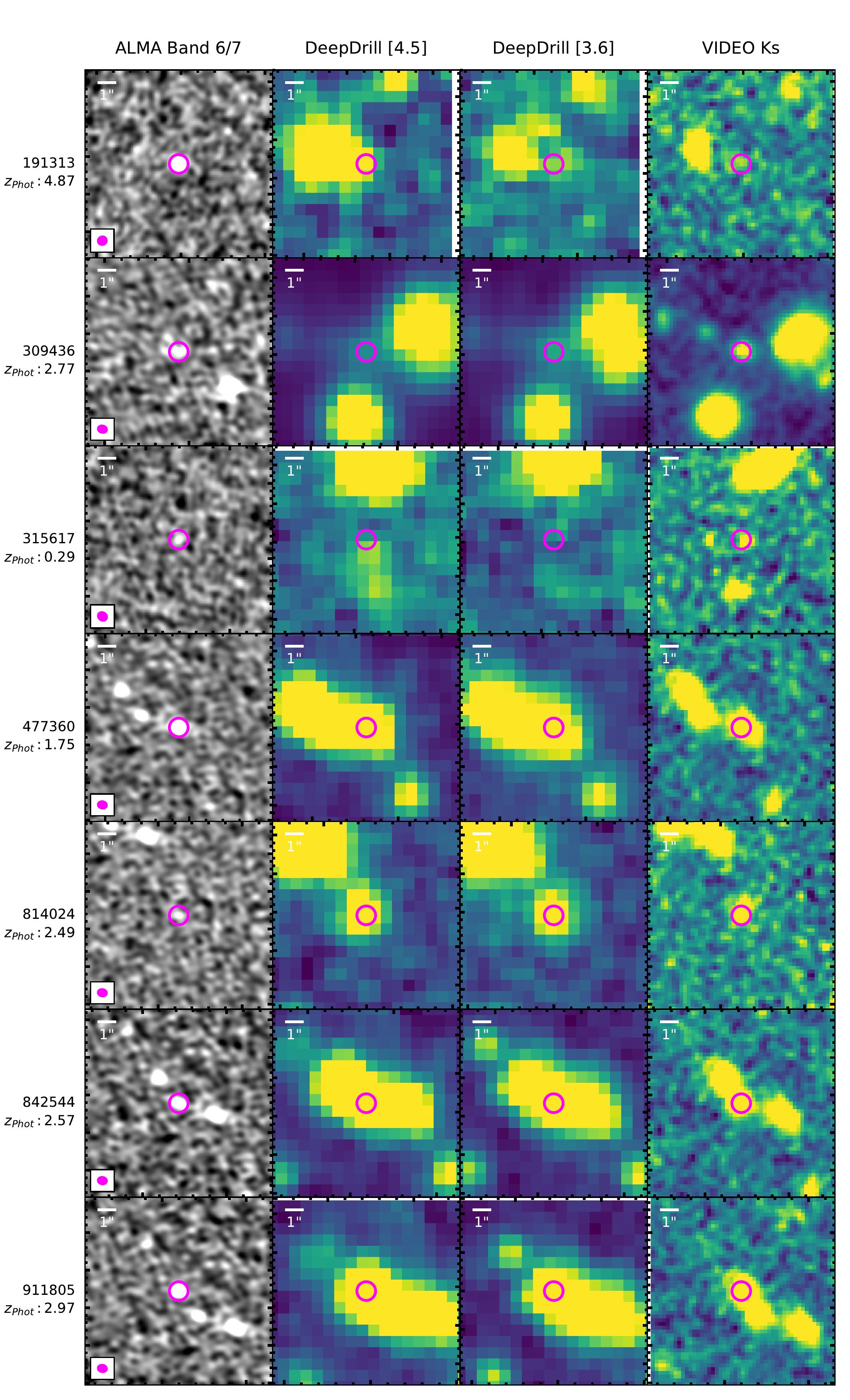}
\caption{Image snapshots of our entire sample in four bands. The grayscale image on the left is the ALMA Band 6/7 image obtained from the archive. The rest are the \textit{Spitzer} 4.5~$\mu$m,  \textit{Spitzer} 3.6~$\mu$m, and VIDEO K$_s$ image left to right. The magenta circle in each image indicates the position of the ALMA source. Our photometric redshift estimates are also given on the left. Each cutout has a size of 10$\arcsec \times$  10$\arcsec$}.\label{fig:cutouts}
\renewcommand{\thefigure}{\arabic{figure} (Cont.)}
\addtocounter{figure}{-1}
\end{figure*}
\begin{figure*}
\centering
\includegraphics[height=0.9\textheight]{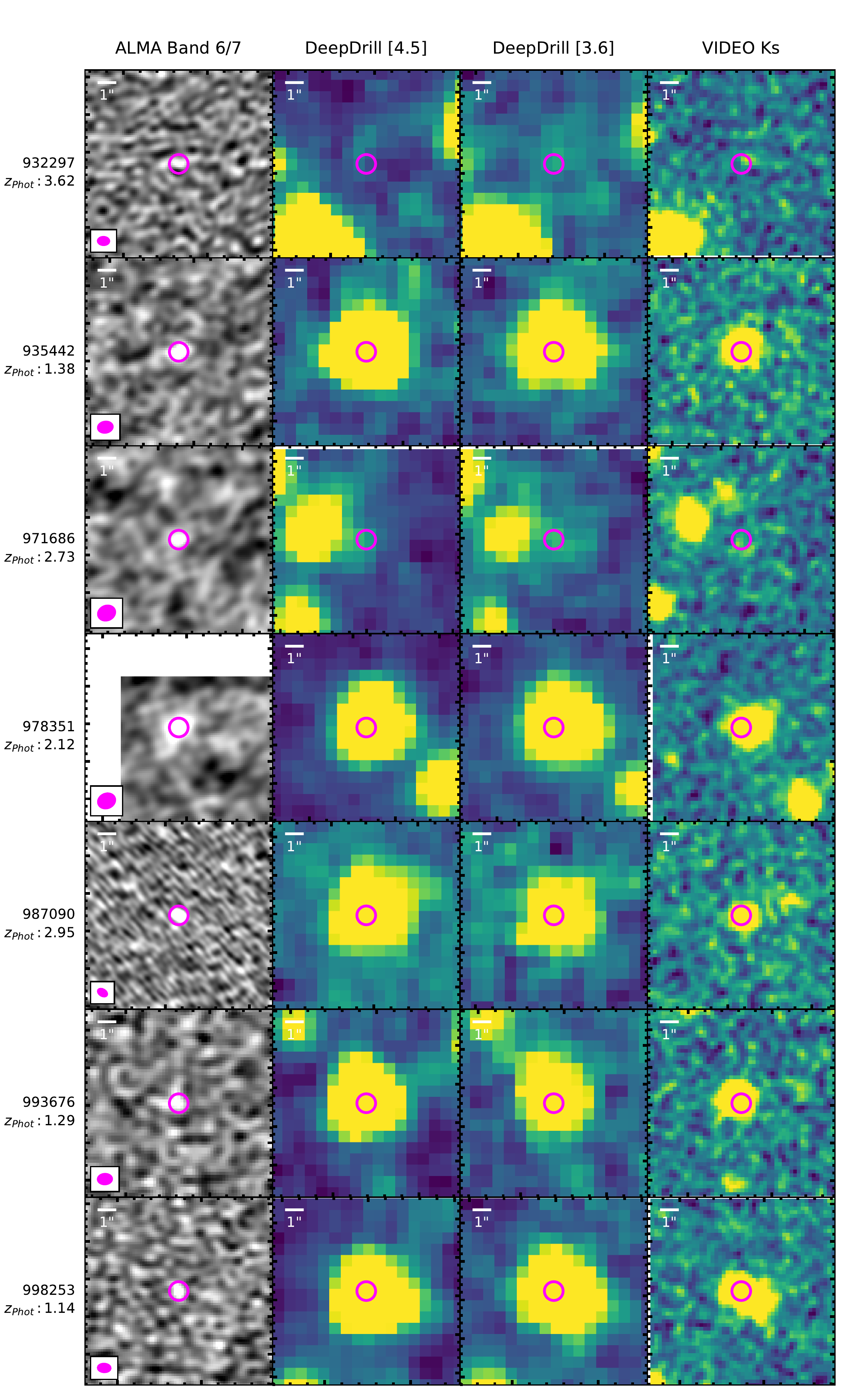}\captcont{\it Continued}
\end{figure*}
\begin{figure*}[htpb!]
\centering
\includegraphics[height=0.9\textheight]{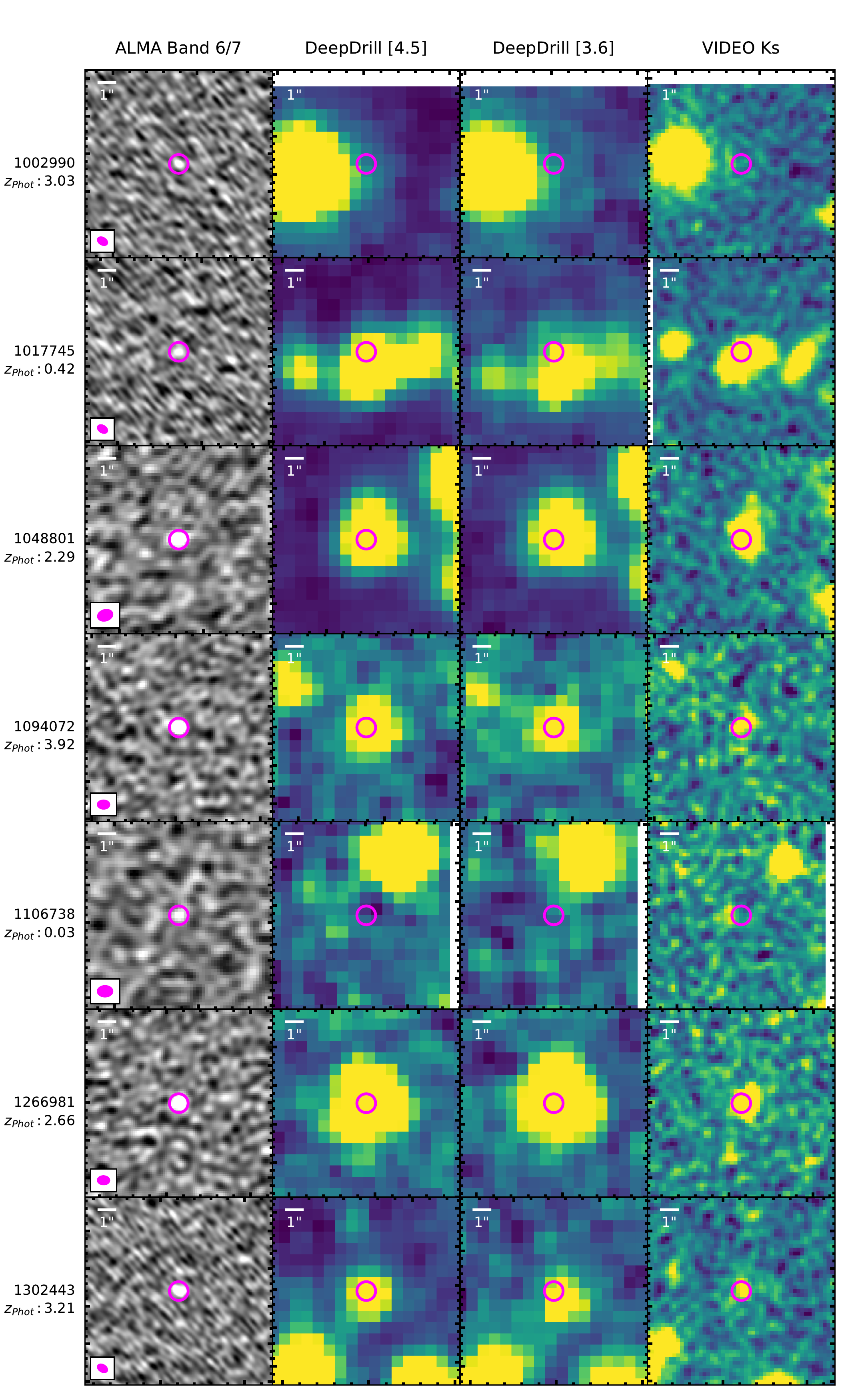}\captcont{{\it Continued}} 
\end{figure*}
\begin{figure*}[htpb!]
\centering
\includegraphics[height=0.9\textheight]{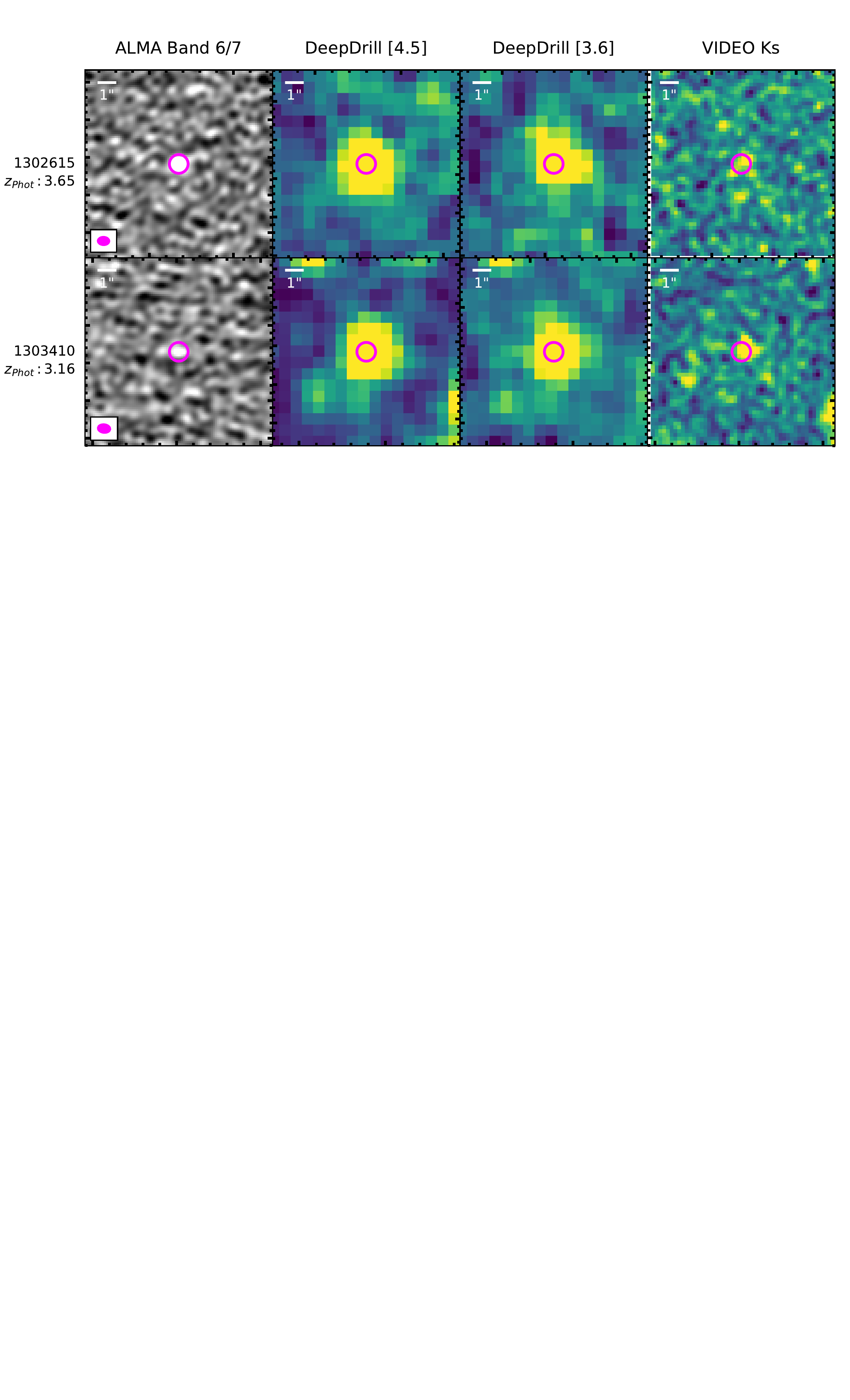}\captcont{\it Continued}
\end{figure*}


\section{The ALMA Archival Projects}\label{sec:appB}
Nine ALMA projects satisfied our selection criteria given above and were publicly available. These projects were undertaken for entirely different science goals. The data is taken from \citet{Ouchi+2013}, who observed a giant starburst galaxy at redshift $z\sim$7, called Himiko, at Band 6. The total integration time was about 3 hours reaching noise levels $\sim$ 19~$\mu$Jy beam$^{-1}$. Other deep maps were taken from ALMA 2013.1.00815.S \citep{Willott+2015}. They have investigated continuum dust emission and [CII] line detections for three more UV-luminous galaxies similar to Himiko found at $z$ $\sim$ 6. The total time on the source is about an hour and a half with rms noise levels of 18~$\mu$Jy beam$^{-1}$. The ALMA 2012.1.00934.S [PI: Phillip Best] covers four maps in the XMM-field to study the star formation activity in moderately star-forming galaxies at $z$ $\sim$ 2.53 using CO molecular line emission. \citet{Ikarashi+2015} have observed 30 potential high-$z$ SMG candidates selected from the AzTEC survey. These galaxies are highly likely to be at $z$$>$3. This project, ALMA 2012.1.00326.S, has 4 minutes on source time with rms levels between 135-65~$\mu$Jy beam$^{-1}$. A 100-minute observation was undertaken by \citet{Inoue+2016} [ALMA 2012.1.00374.S, PI: Kazuaki Ota] to study the state of the epoch of reionization and star formation activity at  $z\sim$7 using spectroscopically confirmed galaxies at that redshift. The same target was observed for another 75 minutes in the project ALMA 2013.A.00021.S to improve the sensitivity of [CII] line detection. 
The data from ALMA 2011.0.00648.S is taken to study interstellar medium of 20 star-forming galaxies at $z\sim$1.4 \citep{Seko+2016a,Seko+2016b}. The typical rms noise levels range from 60$\sim$100~$\mu$Jy beam$^{-1}$ with 10 minutes of the on-source time. \citet{Pentericci+2016} studied a very high redshift galaxy ($z\sim$7) to understand the role of high-$z$ galaxies in the epoch of reionization. The target was observed for 35 minutes which resulted in 20$\,\mu Jy$ noise level. The project 2011.0.00539.S was used to study the ALMA properties of the lensed, bright submillimeter galaxies selected from \textit{Herschel}. The time on each map is about a minute, with typical rms noise levels of 250~$\mu$Jy beam$^{-1}$. This ALMA project is the only Band 7 observations we have used in our search.

\section{Photometry Consistency Check } \label{sec:pcheck}
As we compare colors of the faint SMGs with the selection criteria specified by \citet{Chen+OIR}, it is essential to perform a consistency check within both the data catalogs. Our \textit{Tractor} method utilizes VIDEO catalog whose photometry is in the AB magnitude system. The sample of \citet{Chen+OIR} is taken from UKIDSS-Ultra Deep Survey (UDS) data release 8 (DR8).  We compare $J, H,$ and $K$ bands from the publicly available DR8\footnote{We used the  following link to access the UDS photometry. \url{ http://wsa.roe.ac.uk:8080/wsa/crossID_form.jsp}} with the VIDEO catalog in Figure~\ref{fig:phot_comp}. The data points plotted here are the field galaxies lying within the ALMA pointings used in our search. The positions of those objects were cross-matched with the publicly available UDS DR8 catalog. We show Petrosian magnitudes for both the catalogs in the AB system. Overall, there is a good level of agreement between our photometry and the UDS photometry. We can see a small offset between the $K$ filter used in the UKIDSS-UDS surveys and the $K_s$ filter used in the VIDEO survey. The offset is mainly due to the differences in the filter response functions of  $K$ and $K_s$.

\begin{figure*}[htp]
\renewcommand{\thefigure}{\arabic{figure}}
\addtocounter{figure}{+1}
\includegraphics[width=\textwidth]{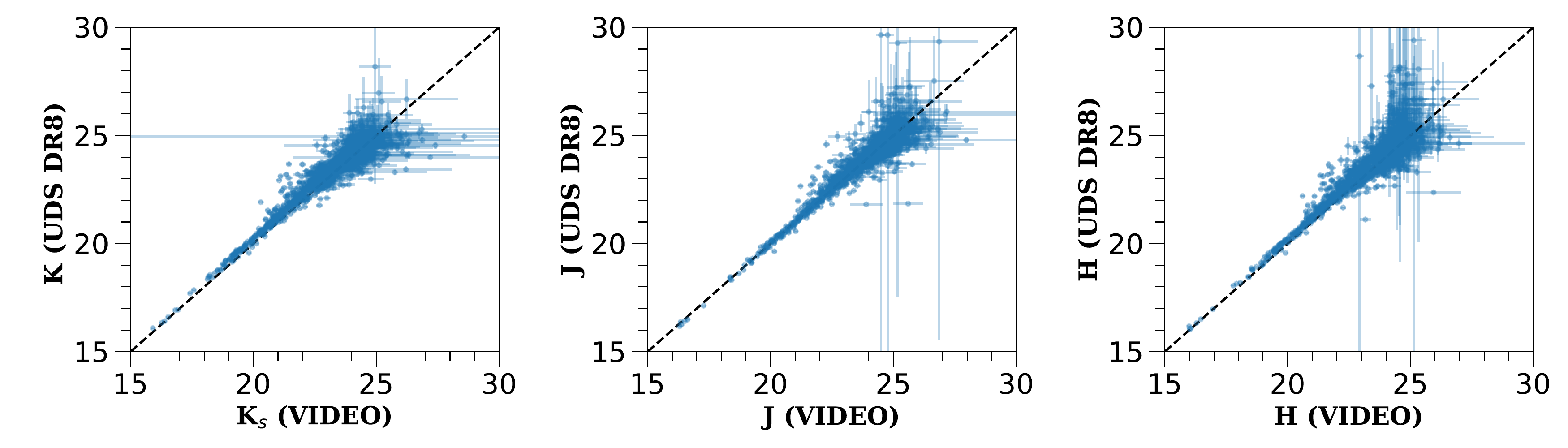}
\caption{The plot shows a comparison of the photometry from our work and the publicly available UKIDSS-UDS DR8 catalog. Our photometry is derived from the VIDEO catalog (See Section~\ref{tractor}). We show Petrosian magnitudes in the AB system for both catalogs. The black dashed line is the equality line. We see overall good agreement between both  the observations.  A slight offset in the $K$ band is mainly due to the differences in the filter response function.\label{fig:phot_comp} }

\end{figure*}

\section{List of Continuum Maps from the ALMA Archive}\label{sec:appA}
Table~\ref{tab:tab1} provides a list of continuum maps used in the search of the faint SMG population. There are 75 individual pointings used in our work, and the sky position, date of observation, project ID, central frequency of observation, clean beam size, and rms noise levels are given in the paper. 

\begin{longrotatetable}
\begin{deluxetable}{cccccccccc}
\tablecaption{Details of the archival observations and individual pointings \label{tab:tab1}}
\tabletypesize\scriptsize
\tablewidth{-2pt}
\tablehead{
\colhead{Index}           & \colhead{Map}      &
\colhead{RA}          & \colhead{Dec}  &
\colhead{Date of Observation}          & \colhead{Project ID}    &
\colhead{$\nu_{obs}$}  & \colhead{$\lambda_{obs}$ }  &
\colhead{Resolving Beam} & \colhead{RMS Noise} \\
& &(J2000) &(J2000) &($yyyy-mm-dd$) & &( GHz)& (mm)  & ($^{\prime\prime} \times ^{\prime\prime} $) & ($\mu$Jy beam$^{-1}$) }
\startdata
0 & CLM1 & 02h28m02.970s & -04d16m18.30s & 2014-06-16 & 2013.1.00815.S & 258 & 1.16 & 0.53 $\times$ 0.46 & 15.42 \\
1 & HiZELS-UDS-NBK-1147 & 02h17m37.000s & -05d09m12.00s & 2014-01-27 & 2012.1.00934.S & 222 & 1.35 & 1.64 $\times$ 0.7 & 58.27 \\
2 & HiZELS-UDS-NBK-1196 & 02h17m40.120s & -05d12m02.35s & 2014-01-27 & 2012.1.00934.S & 222 & 1.35 & 1.65 $\times$ 0.7 & 60.39 \\
3 & HiZELS-UDS-NBK-1348 & 02h17m51.470s & -05d10m36.40s & 2014-01-27 & 2012.1.00934.S & 222 & 1.35 & 1.65 $\times$ 0.7 & 59.21 \\
4 & HiZELS-UDS-NBK-8806 & 02h17m18.650s & -05d07m54.20s & 2014-01-27 & 2012.1.00934.S & 222 & 1.35 & 1.64 $\times$ 0.7 & 59.53 \\
5 & Himiko & 02h17m57.563s & -05d08m44.45s & 2012-07-15 & 2011.0.00115.S & 259 & 1.16 & 0.71 $\times$ 0.48 & 19.86 \\
6 & SXDF1100.013 & 02h16m46.210s & -05d03m47.83s & 2013-11-19 & 2012.1.00326.S & 265 & 1.13 & 0.57 $\times$ 0.38 & 81.4 \\
7 & SXDF1100.019 & 02h17m15.920s & -05d04m03.13s & 2013-11-19 & 2012.1.00326.S & 265 & 1.13 & 0.57 $\times$ 0.38 & 81.23 \\
8 & SXDF1100.027 & 02h17m20.550s & -05d08m42.36s & 2013-11-19 & 2012.1.00326.S & 265 & 1.13 & 0.57 $\times$ 0.38 & 80.98 \\
9 & SXDF1100.036 & 02h18m00.700s & -05d07m29.69s & 2013-11-19 & 2012.1.00326.S & 265 & 1.13 & 0.57 $\times$ 0.38 & 81.5 \\
10 & SXDF1100.039 & 02h18m30.590s & -05d01m16.09s & 2013-11-19 & 2012.1.00326.S & 265 & 1.13 & 0.57 $\times$ 0.38 & 79.86 \\
11 & SXDF1100.045 & 02h18m15.350s & -04d54m03.36s & 2013-11-19 & 2012.1.00326.S & 265 & 1.13 & 0.57 $\times$ 0.38 & 79.46 \\
12 & SXDF1100.049 & 02h17m33.310s & -04d57m02.02s & 2013-11-19 & 2012.1.00326.S & 265 & 1.13 & 0.57 $\times$ 0.38 & 81.66 \\
13 & SXDF1100.053 & 02h16m48.170s & -04d58m56.66s & 2013-11-19 & 2012.1.00326.S & 265 & 1.13 & 0.57 $\times$ 0.38 & 82.26 \\
14 & SXDF1100.060 & 02h17m37.700s & -05d08m23.17s & 2013-11-30 & 2012.1.00326.S & 265 & 1.13 & 0.68 $\times$ 0.48 & 137.39 \\
15 & SXDF1100.063 & 02h17m35.990s & -04d52m18.15s & 2013-11-19 & 2012.1.00326.S & 265 & 1.13 & 0.57 $\times$ 0.38 & 80.8 \\
16 & SXDF1100.073 & 02h18m10.270s & -05d11m26.18s & 2013-11-30 & 2012.1.00326.S & 265 & 1.13 & 0.68 $\times$ 0.48 & 141.14 \\
17 & SXDF1100.082 & 02h17m58.290s & -04d59m11.16s & 2013-11-30 & 2012.1.00326.S & 265 & 1.13 & 0.69 $\times$ 0.48 & 141.18 \\
18 & SXDF1100.083 & 02h17m11.830s & -05d03m59.71s & 2013-11-30 & 2012.1.00326.S & 265 & 1.13 & 0.69 $\times$ 0.48 & 143.26 \\
19 & SXDF1100.090 & 02h17m23.660s & -04d57m24.06s & 2013-11-30 & 2012.1.00326.S & 265 & 1.13 & 0.69 $\times$ 0.48 & 139.67 \\
20 & SXDF1100.101 & 02h18m35.580s & -05d10m02.67s & 2013-11-30 & 2012.1.00326.S & 265 & 1.13 & 0.69 $\times$ 0.48 & 139.46 \\
21 & SXDF1100.104 & 02h18m04.930s & -05d08m19.32s & 2013-11-30 & 2012.1.00326.S & 265 & 1.13 & 0.68 $\times$ 0.48 & 141.65 \\
22 & SXDF1100.109 & 02h18m23.720s & -05d08m05.49s & 2013-11-30 & 2012.1.00326.S & 265 & 1.13 & 0.69 $\times$ 0.48 & 140.09 \\
23 & SXDF1100.110 & 02h17m44.200s & -05d04m08.78s & 2013-11-30 & 2012.1.00326.S & 265 & 1.13 & 0.69 $\times$ 0.48 & 141.69 \\
24 & SXDF1100.123 & 02h18m11.930s & -05d14m20.94s & 2013-11-30 & 2012.1.00326.S & 265 & 1.13 & 0.69 $\times$ 0.48 & 138.85 \\
25 & SXDF1100.127 & 02h17m33.080s & -04d48m51.89s & 2013-11-30 & 2012.1.00326.S & 265 & 1.13 & 0.64 $\times$ 0.46 & 65.92 \\
26 & SXDF1100.154 & 02h18m14.490s & -04d56m12.80s & 2013-11-30 & 2012.1.00326.S & 265 & 1.13 & 0.64 $\times$ 0.46 & 63.9 \\
27 & SXDF1100.174 & 02h18m16.350s & -05d15m28.80s & 2013-11-30 & 2012.1.00326.S & 265 & 1.13 & 0.64 $\times$ 0.46 & 64.95 \\
28 & SXDF1100.179 & 02h18m43.870s & -04d57m33.62s & 2013-11-30 & 2012.1.00326.S & 265 & 1.13 & 0.64 $\times$ 0.46 & 64.43 \\
29 & SXDF1100.230 & 02h17m58.610s & -04d45m49.13s & 2013-11-30 & 2012.1.00326.S & 265 & 1.13 & 0.64 $\times$ 0.46 & 64.6 \\
30 & SXDF1100.231 & 02h17m59.910s & -04d46m50.05s & 2013-11-30 & 2012.1.00326.S & 265 & 1.13 & 0.64 $\times$ 0.46 & 65.54 \\
31 & SXDF1100.233 & 02h18m53.030s & -04d58m14.74s & 2013-11-19 & 2012.1.00326.S & 265 & 1.13 & 0.57 $\times$ 0.38 & 81.82 \\
32 & SXDF1100.250 & 02h17m52.690s & -05d20m31.72s & 2013-11-30 & 2012.1.00326.S & 265 & 1.13 & 0.64 $\times$ 0.46 & 64.78 \\
33 & SXDF1100.253 & 02h18m26.970s & -05d14m38.61s & 2013-11-30 & 2012.1.00326.S & 265 & 1.13 & 0.64 $\times$ 0.46 & 64.45 \\
34 & SXDF1100.276 & 02h18m33.930s & -04d54m23.85s & 2013-11-30 & 2012.1.00326.S & 265 & 1.13 & 0.64 $\times$ 0.46 & 66.35 \\
35 & SXDF1100.277 & 02h17m24.090s & -05d20m21.94s & 2013-11-30 & 2012.1.00326.S & 265 & 1.13 & 0.64 $\times$ 0.46 & 63.16 \\
36 & SXDF.220GHZ & 02h18m56.536s & -05d19m58.92s & 2014-05-03 & 2012.1.00374.S & 224 & 1.33 & 0.82 $\times$ 0.62 & 12.56 \\
37 & SXDF-NB1 & 02h18m56.536s & -05d19m58.92s & 2015-12-27 & 2013.A.00021.S & 224 & 1.33 & 1.94 $\times$ 1.07 & 15.1 \\
38 & SXDS1\_13015 & 02h17m13.617s & -05d09m39.82s & 2012-08-15 & 2011.0.00648.S & 243 & 1.23 & 0.8 $\times$ 0.58 & 100.87 \\
39 & SXDS1\_1723 & 02h17m32.701s & -05d13m16.47s & 2012-08-26 & 2011.0.00648.S & 240 & 1.25 & 1.11 $\times$ 0.59 & 111.28 \\
40 & SXDS1\_33244 & 02h16m47.397s & -05d03m28.09s & 2012-08-26 & 2011.0.00648.S & 240 & 1.25 & 0.95 $\times$ 0.59 & 77.72 \\
41 & SXDS1\_35572 & 02h17m34.652s & -05d02m38.97s & 2012-08-11 & 2011.0.00648.S & 236 & 1.27 & 0.78 $\times$ 0.58 & 46.17 \\
42 & SXDS1\_59863 & 02h17m45.877s & -04d54m37.60s & 2012-08-15 & 2011.0.00648.S & 243 & 1.23 & 0.81 $\times$ 0.58 & 95.02 \\
43 & SXDS1\_59914 & 02h17m12.977s & -04d54m40.40s & 2012-08-26 & 2011.0.00648.S & 240 & 1.25 & 1.04 $\times$ 0.59 & 109.0 \\
44 & SXDS1\_67002 & 02h19m02.654s & -04d49m55.83s & 2012-08-15 & 2011.0.00648.S & 243 & 1.23 & 0.83 $\times$ 0.57 & 97.8 \\
45 & SXDS1\_68849 & 02h17m00.276s & -04d48m14.53s & 2012-08-26 & 2011.0.00648.S & 240 & 1.25 & 0.85 $\times$ 0.6 & 96.21 \\
46 & SXDS1\_79518 & 02h18m59.059s & -04d51m24.91s & 2012-08-26 & 2011.0.00648.S & 240 & 1.25 & 0.9 $\times$ 0.6 & 99.08 \\
47 & SXDS1 & 02h17m05.789s & -04d51m25.68s & 2012-08-09 & 2011.0.00648.S & 231 & 1.3 & 0.8 $\times$ 0.67 & 64.28 \\
48 & SXDS2\_13316 & 02h17m39.035s & -04d44m41.75s & 2012-08-15 & 2011.0.00648.S & 243 & 1.23 & 0.81 $\times$ 0.57 & 92.28 \\
49 & SXDS2\_22198 & 02h17m53.416s & -04d42m53.42s & 2012-08-11 & 2011.0.00648.S & 236 & 1.27 & 0.81 $\times$ 0.58 & 44.84 \\
50 & SXDS2 & 02h17m24.356s & -05d00m44.85s & 2012-08-09 & 2011.0.00648.S & 231 & 1.3 & 0.8 $\times$ 0.66 & 63.38 \\
51 & SXDS3\_101746 & 02h18m04.178s & -05d19m38.28s & 2012-08-26 & 2011.0.00648.S & 240 & 1.25 & 0.93 $\times$ 0.6 & 99.25 \\
52 & SXDS3\_103139 & 02h16m57.652s & -05d14m34.86s & 2012-08-11 & 2011.0.00648.S & 236 & 1.27 & 0.77 $\times$ 0.58 & 46.08 \\
53 & SXDS3\_110465 & 02h18m20.953s & -05d19m07.71s & 2012-08-26 & 2011.0.00648.S & 240 & 1.25 & 0.99 $\times$ 0.6 & 99.74 \\
54 & SXDS3 & 02h17m13.683s & -05d04m07.66s & 2012-08-09 & 2011.0.00648.S & 231 & 1.3 & 0.81 $\times$ 0.66 & 60.34 \\
55 & SXDS5\_19723 & 02h16m24.372s & -05d09m18.05s & 2012-08-11 & 2011.0.00648.S & 236 & 1.27 & 0.77 $\times$ 0.58 & 45.41 \\
56 & SXDS5\_28019 & 02h16m08.532s & -05d06m15.61s & 2012-08-11 & 2011.0.00648.S & 236 & 1.27 & 0.78 $\times$ 0.58 & 46.36 \\
57 & SXDS5\_9364 & 02h16m33.807s & -05d13m44.68s & 2012-08-15 & 2011.0.00648.S & 243 & 1.23 & 0.81 $\times$ 0.57 & 97.7 \\
58 & UDS16 & 02h17m25.474s & -05d11m08.25s & 2016-03-31 & 2015.1.01105.S & 241 & 1.24 & 0.96 $\times$ 0.8 & 21.5 \\
59 & WMH5 & 02h26m27.030s & -04d52m38.30s & 2014-06-16 & 2013.1.00815.S & 262 & 1.14 & 0.5 $\times$ 0.48 & 18.05 \\
60 & XMMF10 & 02h16m31.850s & -05d53m22.20s & 2012-06-17 & 2011.0.00539.S & 343 & 0.87 & 0.51 $\times$ 0.41 & 220.03 \\
61 & XMMF11 & 02h21m35.220s & -06d26m16.97s & 2012-06-17 & 2011.0.00539.S & 343 & 0.87 & 0.51 $\times$ 0.41 & 261.51 \\
62 & XMMF12 & 02h22m50.830s & -03d24m13.81s & 2012-06-17 & 2011.0.00539.S & 342 & 0.87 & 0.49 $\times$ 0.41 & 212.05 \\
63 & XMMF13 & 02h18m41.550s & -03d50m01.90s & 2012-06-17 & 2011.0.00539.S & 343 & 0.87 & 0.51 $\times$ 0.41 & 227.81 \\
64 & XMMF14 & 02h20m29.140s & -06d48m45.89s & 2012-06-17 & 2011.0.00539.S & 343 & 0.87 & 0.51 $\times$ 0.41 & 236.57 \\
65 & XMMF15 & 02h22m05.540s & -07d07m27.20s & 2012-06-17 & 2011.0.00539.S & 343 & 0.87 & 0.51 $\times$ 0.41 & 212.96 \\
66 & XMMF16 & 02h19m42.910s & -05d24m32.54s & 2012-06-17 & 2011.0.00539.S & 343 & 0.87 & 0.51 $\times$ 0.41 & 224.83 \\
67 & XMMF17 & 02h19m18.400s & -03d10m51.30s & 2012-06-17 & 2011.0.00539.S & 343 & 0.87 & 0.51 $\times$ 0.42 & 259.53 \\
68 & XMMF18 & 02h29m44.790s & -03d41m10.17s & 2012-06-17 & 2011.0.00539.S & 343 & 0.87 & 0.52 $\times$ 0.41 & 300.12 \\
69 & XMMF19 & 02h20m21.610s & -01d53m29.12s & 2012-06-17 & 2011.0.00539.S & 343 & 0.87 & 0.52 $\times$ 0.42 & 277.51 \\
70 & XMMF4 & 02h20m16.689s & -06d01m44.38s & 2012-06-17 & 2011.0.00539.S & 343 & 0.87 & 0.51 $\times$ 0.41 & 262.64 \\
71 & XMMF5 & 02h22m01.576s & -03d33m40.60s & 2012-06-17 & 2011.0.00539.S & 343 & 0.87 & 0.52 $\times$ 0.41 & 233.23 \\
72 & XMMF6 & 02h25m48.210s & -04d17m51.10s & 2012-06-17 & 2011.0.00539.S & 343 & 0.87 & 0.51 $\times$ 0.41 & 254.89 \\
73 & XMMF7 & 02h18m53.100s & -06d33m23.70s & 2012-06-17 & 2011.0.00539.S & 343 & 0.87 & 0.51 $\times$ 0.41 & 219.54 \\
74 & XMMF8 & 02h30m05.980s & -03d41m52.70s & 2012-06-17 & 2011.0.00539.S & 343 & 0.87 & 0.52 $\times$ 0.41 & 254.45 \\
\enddata
\end{deluxetable}
\end{longrotatetable}

\begin{figure*}
\centering
\includegraphics[trim={1.1cm 0 1cm 0},clip, height=0.85\textheight]{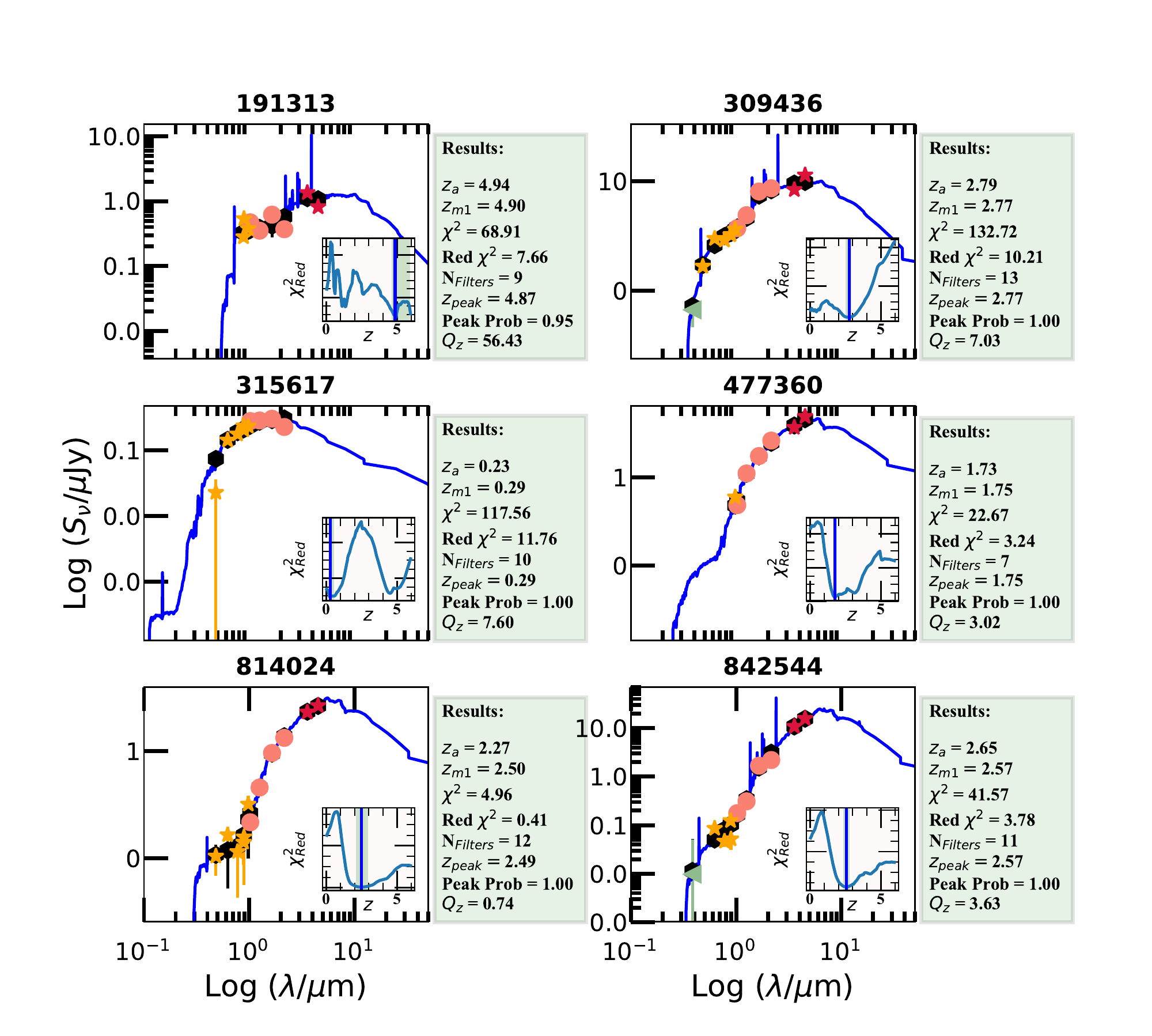}
\caption{The figure sets show photometric redshift fitting of the entire sample. Different colored symbols correspond to the observed 13-band photometry, which is obtained using the {\it Tractor} image modeling code (Section~\ref{tractor}). Red stars highlight {\it Spitzer} DeepDrill measurements at 3.6 and 4.5~$\mu$m, orange circles denote measurements from the VIDEO survey,  the green triangle represents the optical $u$-band data from CFHTLS, and the yellow stars mark the optical data points from HSC. The Blue solid line is the actual SED fit, and the black filled circles show the estimated template flux for each filter. The green box on the right lists key output parameters from EAZY. The inset figure shows the $\chi^{2}$ distribution of the fit at each redshift bin. The blue vertical line shows the peak of the redshift distribution after applying the prior, and the light green regions are the 95\% confidence intervals of the fit. {\label{fig:eazy_all}}}
\end{figure*}

\begin{figure*}
\centering
\renewcommand{\thefigure}{\arabic{figure}}
\addtocounter{figure}{-1}
\includegraphics[trim={1.1cm 0 1cm 0},clip, height=0.85\textheight]{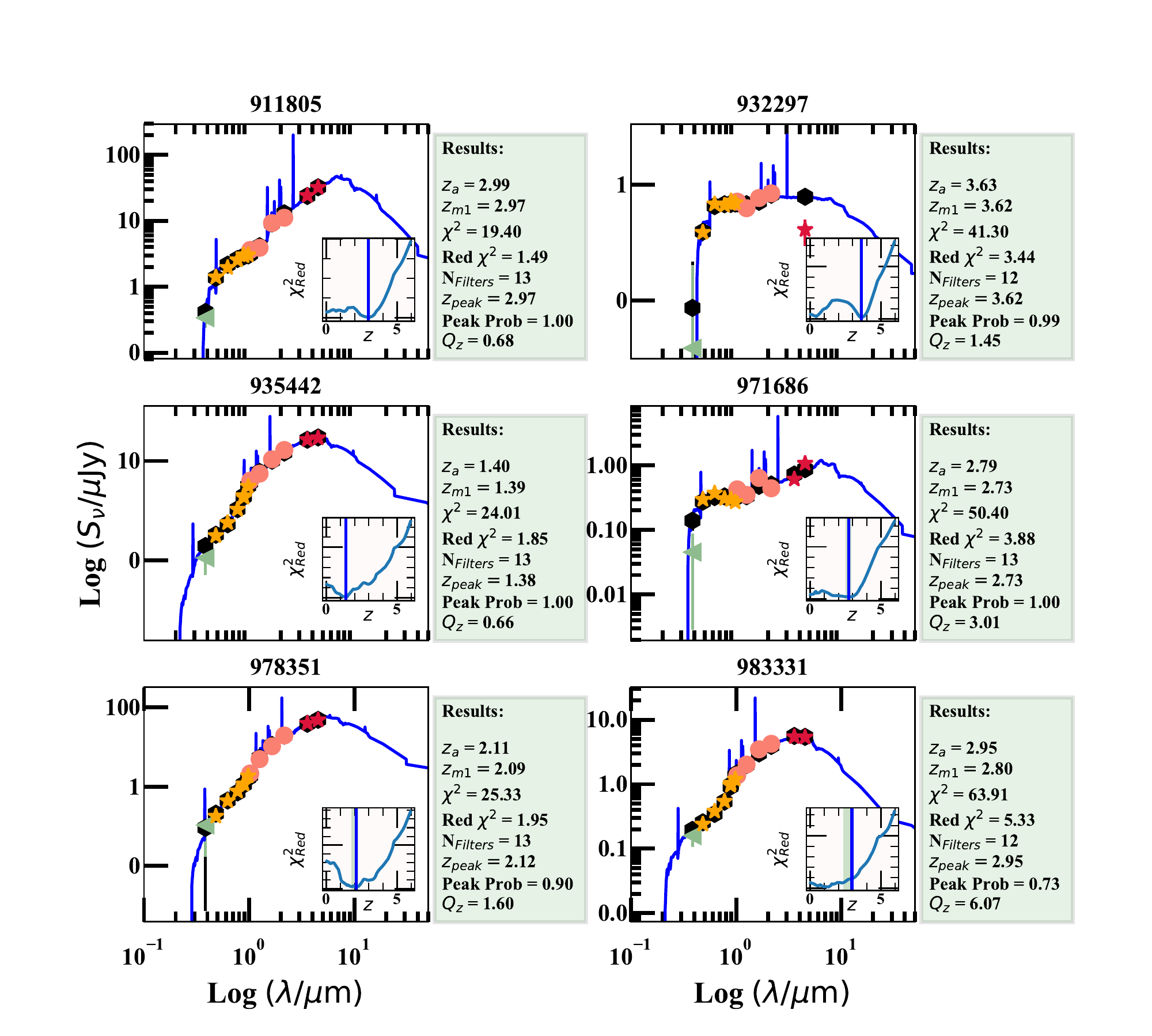}
\captcont{Continued}
\end{figure*}

\begin{figure*}
\centering

\includegraphics[trim={1.1cm 0 1cm 0},clip, height=0.85\textheight]{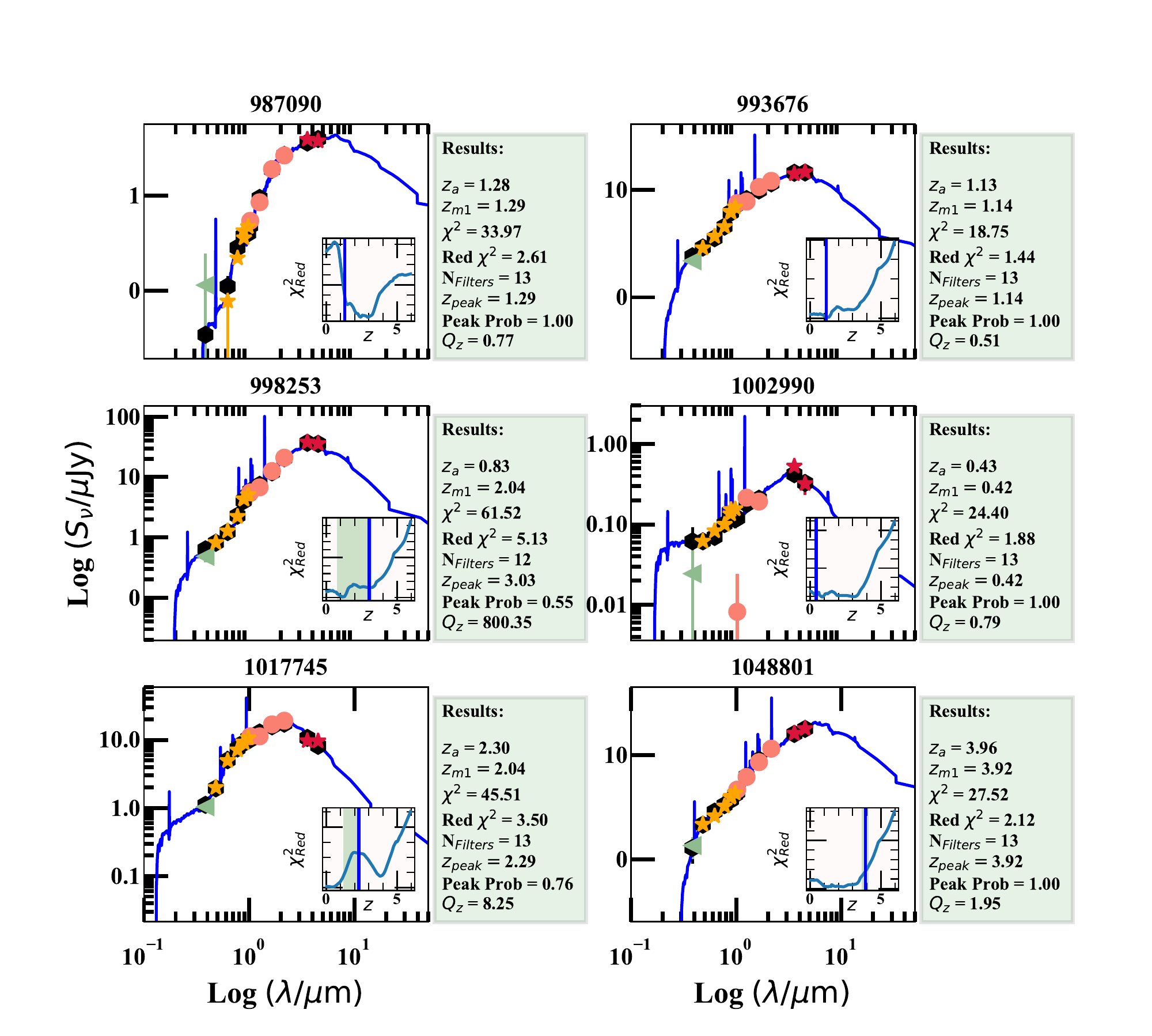}
\captcont{Continued}
\end{figure*}

\begin{figure*}
\centering
\includegraphics[trim={1.1cm 0 1cm 0},clip, height=0.85\textheight]{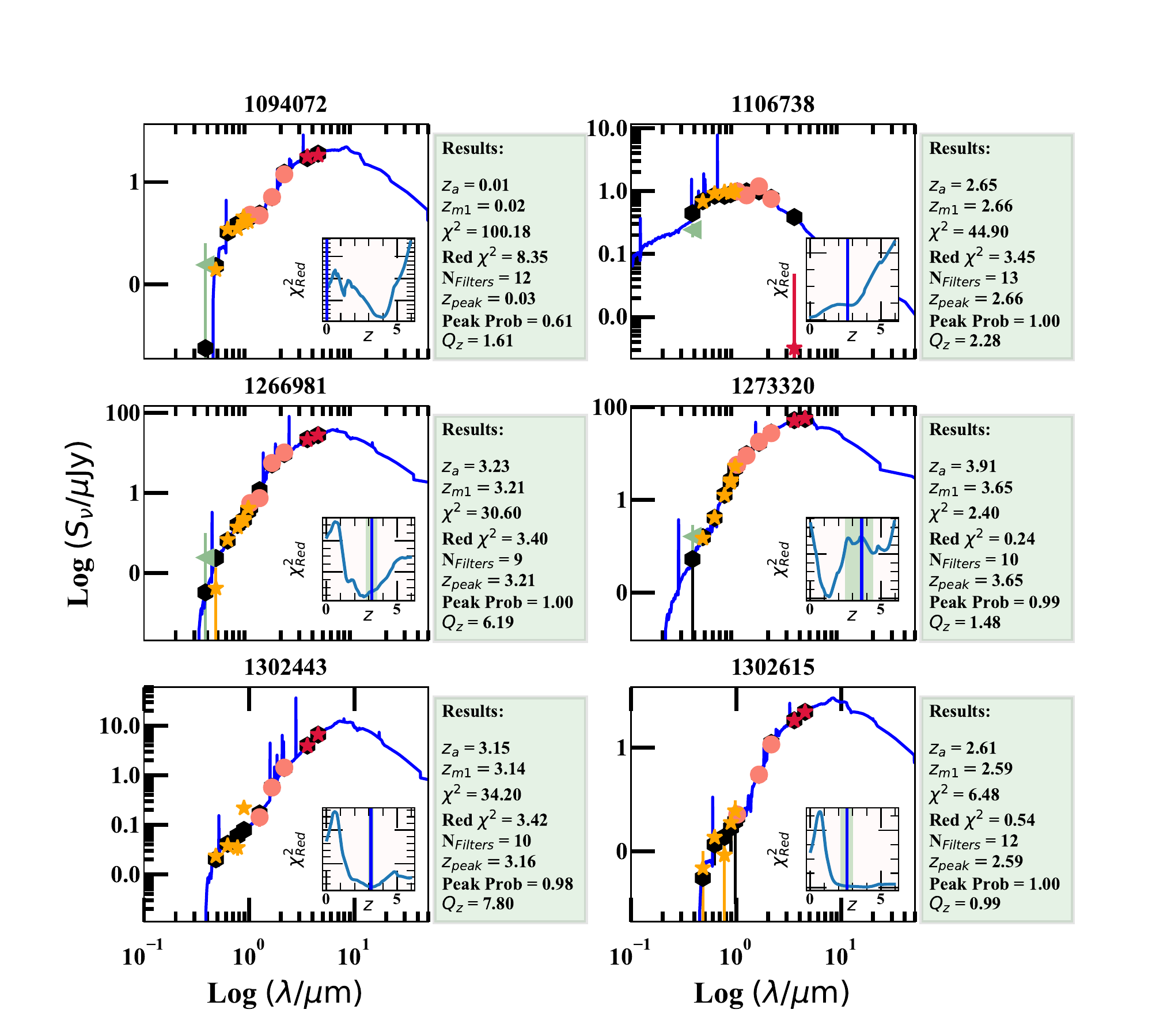}
\captcont{Continued}
\end{figure*}

\begin{figure*}
\centering
\includegraphics[trim={1.1cm 0 1cm 0},clip, height=0.85\textheight]{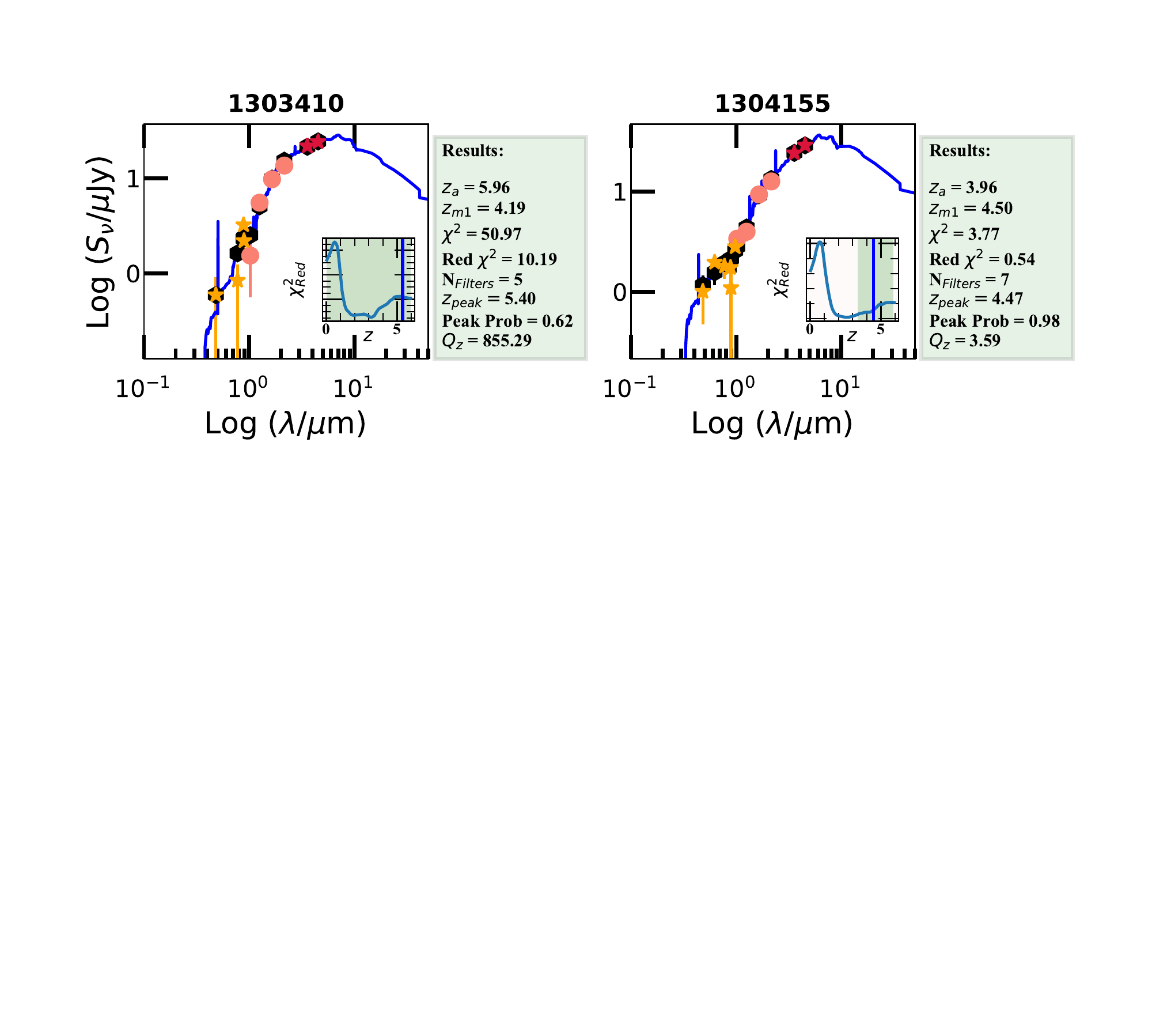}
\captcont{Continued}
\end{figure*}
\clearpage

\bibliographystyle{apj}

\bibliography{ALMA_pallavi_tractor_text}


%

\end{document}